\documentclass[a4paper,11pt]{article}

\pdfoutput=1
\usepackage{jinstpub}

\title{Performance of scintillating tiles with direct silicon-photomultiplier (SiPM) readout for application to large area detectors}

\author[a]{A.~Balla,}
\author[a]{B.~Buonomo,}
\author[b]{V.~Cafaro,}
\author[a]{A.~Calcaterra,}
\author[a]{F.~Cardelli,}
\author[a]{P.~Ciambrone,}
\author[b,c]{V.~Cicero,}
\author[a]{D.~Di~Giovenale,}
\author[a]{C.~Di~Giulio,}
\author[a]{G.~Felici,}
\author[a]{L.G.~Foggetta,}
\author[b]{V.~Giordano,}
\author[a]{G.~Lanfranchi,}
\author[b]{I.~Lax,}
\author[b]{A.~Montanari,}
\author[a]{G.~Papalino,}
\author[a]{A.~Paoloni,}
\author[b,c] {T.~Rovelli,}
\author[a]{A.~Saputi,}
\author[b]{G.~Torromeo,}
\author[b] {N.~Tosi.}

\affiliation[a]{INFN - Laboratori Nazionali di Frascati, via E. Fermi 40, 00044 Frascati (Rome), Italy}
\affiliation[b]{INFN - Sezione di Bologna, Viale Berti Pichat, 6/2, 40127 Bologna, Italy}
\affiliation[c]{Dipartimento di Fisica e Astronomia, Universit\`a di Bologna, Viale Berti Pichat, 6/2, 40127 Bologna, Italy}

\emailAdd{alessandro.paoloni@lnf.infn.it}

\date{September 2021}
\abstract{ The light yield, the time resolution and the efficiency of different types of scintillating tiles with direct Silicon Photomultiplier readout and instrumented with a customised front-end electronics have been measured at the Beam Test Facility of Laboratori Nazionali di Frascati and several test stands.
The results obtained with different configurations are presented.  
A time resolution of the order of 300~ps, a light yield of more than 230 photo-electrons, and an efficiency better than 99.8\% are obtained with $\sim 225$ cm$^2$ large area tiles.
This technology is suitable for a wide range of applications in high-energy physics, in particular for large area muon and timing detectors.}


\arxivnumber{xxxx.yyyy} 

\begin{document}

\maketitle

\section{Introduction}
\label{sec:introduction}

Organic scintillators offer a fast response and high light yield for moderate cost, making them a good choice for the application in large area detectors for particle physics.
The main reasons for choosing organic scintillators are their fast response (short rise and decay times) and their high light yield. 
Silicon Photo-Multipliers (SiPMs)~\footnote{Also known as Multi-Pixel Photon Counters (MPPCs).}) \cite{sipm} provide advantageous properties such as good timing, compactness, and high Photon Detection Efficiency (PDE).
Scintillating tiles with direct SiPM readout, pioneered for application in hadron calorimeters for electron-positron colliders \cite{directsipm} \cite{directsipm2}, allow compact detectors with high granularity and good time resolution to be built.

In addition to a very good timing performance, the choice of tiles guarantees the determination of the $x,y$ coordinates with a single active layer and a good tolerance against hit rate variations. Moreover the construction and assembly procedure is modular and therefore can be easily shared among different production sites, which is paramount for the construction of large area detectors.

This article describes the performance obtained on 225 cm$^2$ area scintillating tiles with direct SiPM readout developed for possible use in large area muon systems, as for example, the muon system of the proposed SHiP experiment \cite{ship}.
However, the modular structure and the very competitive cost render this system suitable to multiple applications in high energy physics and beyond, in particular for large area detectors.

\begin{figure}[hbt]
  \begin{center}
 \includegraphics[width=0.8\textwidth]{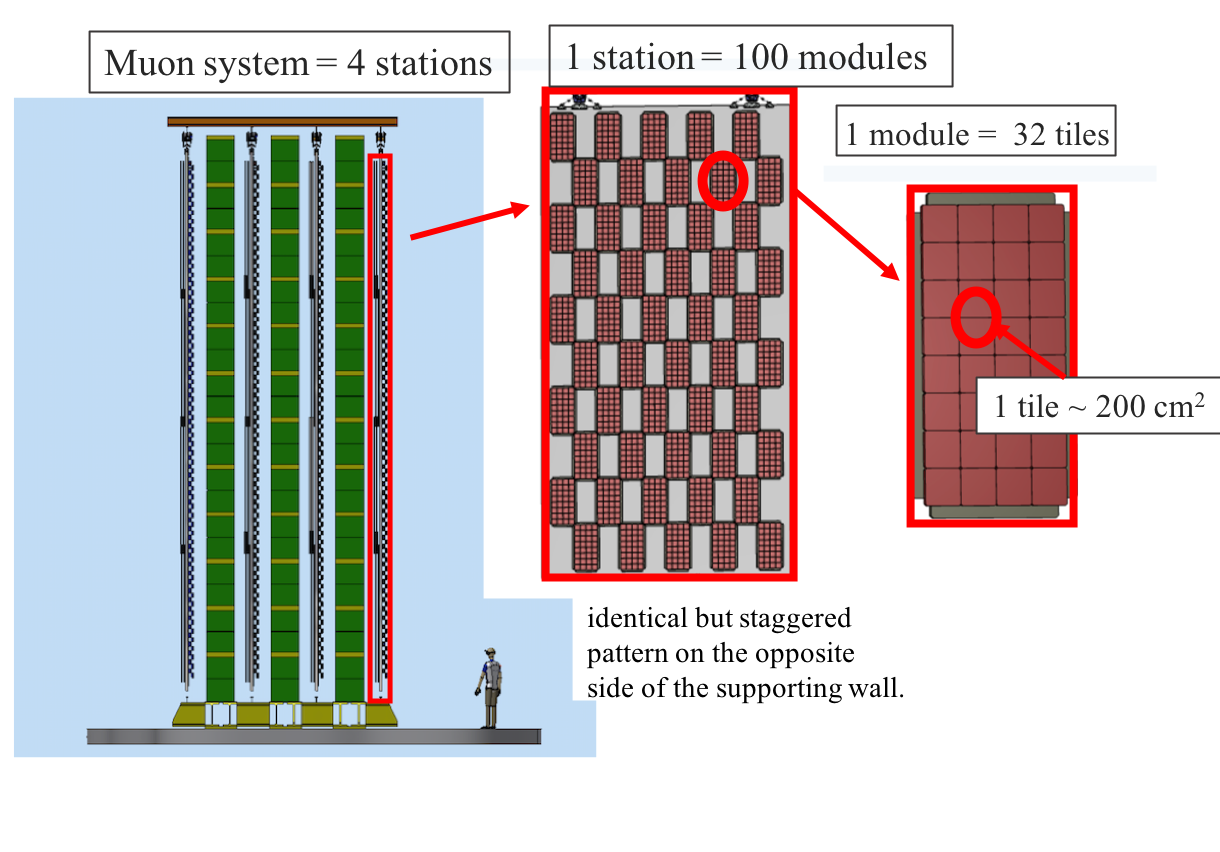}
    \caption{Schematic layout of the SHiP muon system. The system is made of four active stations interleaved by passive filters. Each station is made by 100 modules, 32-tiles each, organised in a chess-like structure. Each tile has an area of $\sim 200$~cm$^{2}$.}
\label{fig:ship-muon}
\end{center}
\end{figure}

An example of a large area muon detector based on scintillating tiles is shown in Figure~\ref{fig:ship-muon}. Each of the four stations is made of 100~modules, 32 tiles each.
The modules are installed on both sides of a supporting structure (Al wall) in a staggered {\it chess} structure:
such a design allows both the dead zones to be minimised and the weight on the supporting wall to be equally shared between the two wall sides. Shape and size of the tiles and the muon stations can be modified depending on the required application.

\section{The prototypes}
\label{sec:prototypes}

Four tile prototypes have been built and characterised at several test stands  in Frascati and Bologna INFN laboratories and their performance assessed during a test beam at the Beam Test Facility (BTF) of Laboratori Nazionali di Frascati in January 2021.

\begin{figure}[hbt]
  \begin{center}
      \includegraphics[width=0.7\textwidth]{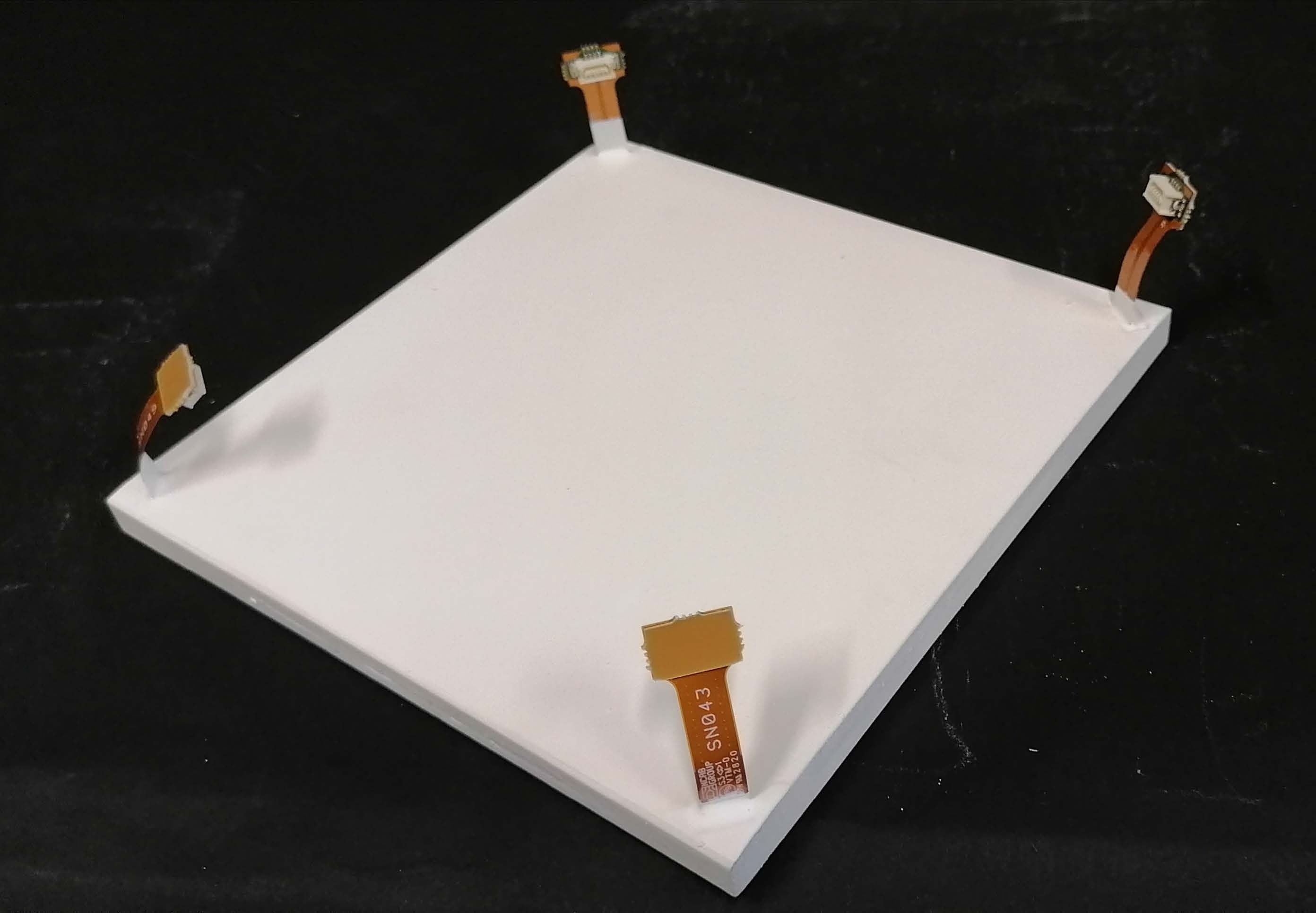}  
    \caption{Tile prototype with white reflecting painting. The four SiPMs are attached to flex cables glued in grooves inside the scintillator at the tile corners.}
\label{fig:tile_white}
\end{center}
\end{figure}

Two (out of four) tiles are made of EJ200 cast organic scintillator from Eljen company~\footnote{https://eljentechnology.com} and the other two by organic scintillator from UNIPLAST (Vladimir, Russia).
The  tiles  dimensions are of ($150 \times 150 \times 10$)~mm$^3$ for both UNIPLAST and for Eljen company. Each tile is read out by four SiPMs placed at the tile corners, either engraved into slots dug in the scintillator or glued at the cut corners.
An example of prototype with the four SiPMs connected to flex cables and glued in slots engraved at the tile corners is shown in Figure~\ref{fig:tile_white}.
Three out of four tiles are painted with three layers of reflective painting 
Eljen EJ-510~\footnote{https://eljentechnology.com/products/accessories/ej-510-ej-520.}, while the fourth one is covered by a chemical reflector obtained by etching the scintillator surface in a chemical agent, that results in the formation of a white micropore deposit over polystyrene \cite{scintillator}.
All tiles are wrapped with Teflon tape to ensure light tightness.
The main characteristics of the four tiles under test are summarised in Table~\ref{tab:tiles}.

\begin{table}[htbp]
\caption{Main characteristics of the four tiles under test and average breakdown voltage of the four SiPMs in each tile.}
\label{tab:tiles}
\vspace{.1cm}
\begin{center}
\begin{tabular}{llllll}
\hline \hline
       & Scintillator & Dimensions & SiPM placement &  coating & $V_{\rm break}$ (V) \\ \hline
Tile 1 &  EJ200        & (150$\times 150 \times 10)$~mm$^3$ & slots & painting + Teflon & 38.34 $\pm$ 0.06 \\
Tile 2 &  EJ200        & (150$\times 150 \times 10)$~mm$^3$ & corners       & painting + Teflon & 38.43 $\pm$ 0.09 \\
Tile 3 &  UNIPLAST & (150$\times 150 \times 10)$~mm$^3$ & slots & etching + Teflon & 38.55 $\pm$ 0.05 \\
Tile 4 &  UNIPLAST & (150$\times 150 \times 10)$~mm$^3$ & slots & painting + Teflon & 38.31 $\pm$ 0.08 \\ 
\hline \hline
\end{tabular}
\end{center}
\end{table}

\vskip 2mm
The tiles are read out by four SiPMs, Hamamatsu S14160-6050HS, with $6\times 6$~mm$^2$ active area.
The other characteristics of the SiPMs, as deduced from the datasheet\footnote{https://www.hamamatsu.com/eu/en/product/type/S14160-6050HS/index.html}, are reported on table \ref{tab:sipmchar}.
The V-I curve  of every SiPM has been measured and the breakdown voltage ($V_{\rm break}$) determined.
SiPMs with similar $V_{\rm break}$ have been grouped on each tile, for which the average $V_{\rm break}$ is also provided on Table~\ref{tab:tiles}, together with the rms of the four values.
The SiPMs are mounted on short Kapton flex cables that ensure the connection with the motherboard (see Section~\ref{sec:fee}). The flex cable end hosting the SiPM is glued to the tile via optical glue Eljen EJ-500\footnote{https://eljentechnology.com/products/accessories/ej-500.}. The other end is connected  to mezzanines placed on the motherboard attached to the tile surface, where the Front-End Electronics (FEE) is located.
All the details of the FEE are given in Section~\ref{sec:fee}. 

\begin{table}[htbp]
\caption{Main characteristics of Hamamatsu S14160-6050HS SiPMs. The PDE, gain, dark current and crosstalk values are given for the suggested operation voltage of $V_{\rm break}$+2.7 V.}
\label{tab:sipmchar}
\vspace{.1cm}
\begin{center}
\begin{tabular}{llllll}
\hline \hline
Number of channels & 1 \\ 
Pixel pitch & 50 $\mu$m \\
Number of pixels & 14331 \\
$V_{\rm break}$ & $\sim$ 38 V \\
PDE & 50\% \\
Gain & o(10$^6$) \\
Typical dark current value & 2.5 $\mu$A \\
Crosstalk probability & 7\% \\
\hline \hline
\end{tabular}
\end{center}
\end{table}

\section{Front-End electronics} 
\label{sec:fee}

\begin{figure}[ht]
  \begin{center}
   \includegraphics[width=0.3\textwidth]{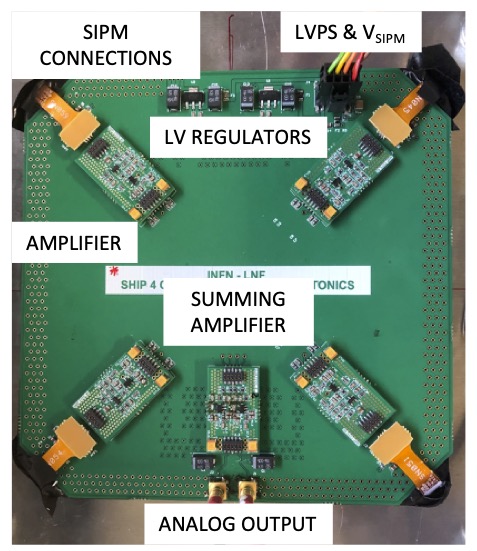}
   \includegraphics[width=0.6\textwidth]{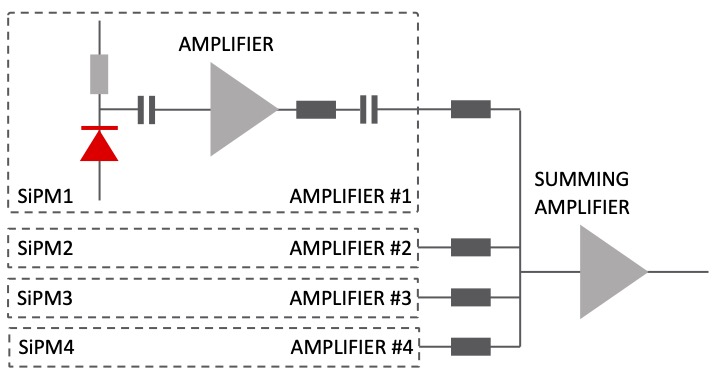}
    \caption{ Picture of a fully instrumented tile (left) and its readout circuit block diagram (right).}
\label{fig:fee_layout}
\end{center}
\end{figure}

The FEE for the tile prototypes has been designed following a mother-board/daughter-boards approach with on-the-shelf components.
In this way different amplifier topologies can be tested without affecting the SiPM-board connections.
The picture of a fully instrumented scintillator tile is shown in Figure ~\ref{fig:fee_layout}.

As well known the time resolution depends on both signal slew rate and noise, therefore photon collection must be maximised by means of large area SiPMs and high bandwidth readout electronics. 
Unfortunately large area SiPMs have a large parasitic capacitance that increases the amplifier input noise, reducing the overall signal-to-noise S/N ratio. A reduction of the parasitic capacitance could be achieved by connecting SiPMs in series, but this configuration requires higher supply voltages and reduces the signal amplitude and therefore it is not the best choice for a topology where SiPMs are spread over several cm$^2$ area.

For our readout topology, SiPMs instrumented with individual preamplifiers followed by a common summing point is a preferable solution as it allows local signal amplification and the possibility of adjusting the shaping time both at the preamplifier input and at the common summing point.

The readout circuit block diagram is shown in Figure \ref{fig:fee_layout}: each SiPM output is amplified and combined with signals from the other amplifiers by means of a summing amplifier. The summing amplifier output is, finally, routed to a digitizer. 
The mother board connections have been designed to equalise the propagation delay of the signals from the four (local) SiPM amplifier. 

The mother-board PCB includes connectors for both preamplifiers and summing amplifier, SiPM bias circuits and low-voltage regulators. Four layers PCBs have been used for proper impedance signal routing traces and low-impedance supply voltage distribution. MCX connectors have been chosen for analog signal transmission while MOLEX nano-fit connectors have been used for low-voltage and SiPM bias distribution.

The SiPMs are connected to the preamplifier via low impedance short kapton flex cables allowing easy replacement of the full readout circuit in case of failures. 

Because of the SiPMs large parasitic capacitance (of the order of 2~nF), front-end preamplifiers with low input impedance must be used.

\begin{figure}[ht]
 \begin{center}
	\includegraphics[width=0.9\textwidth]{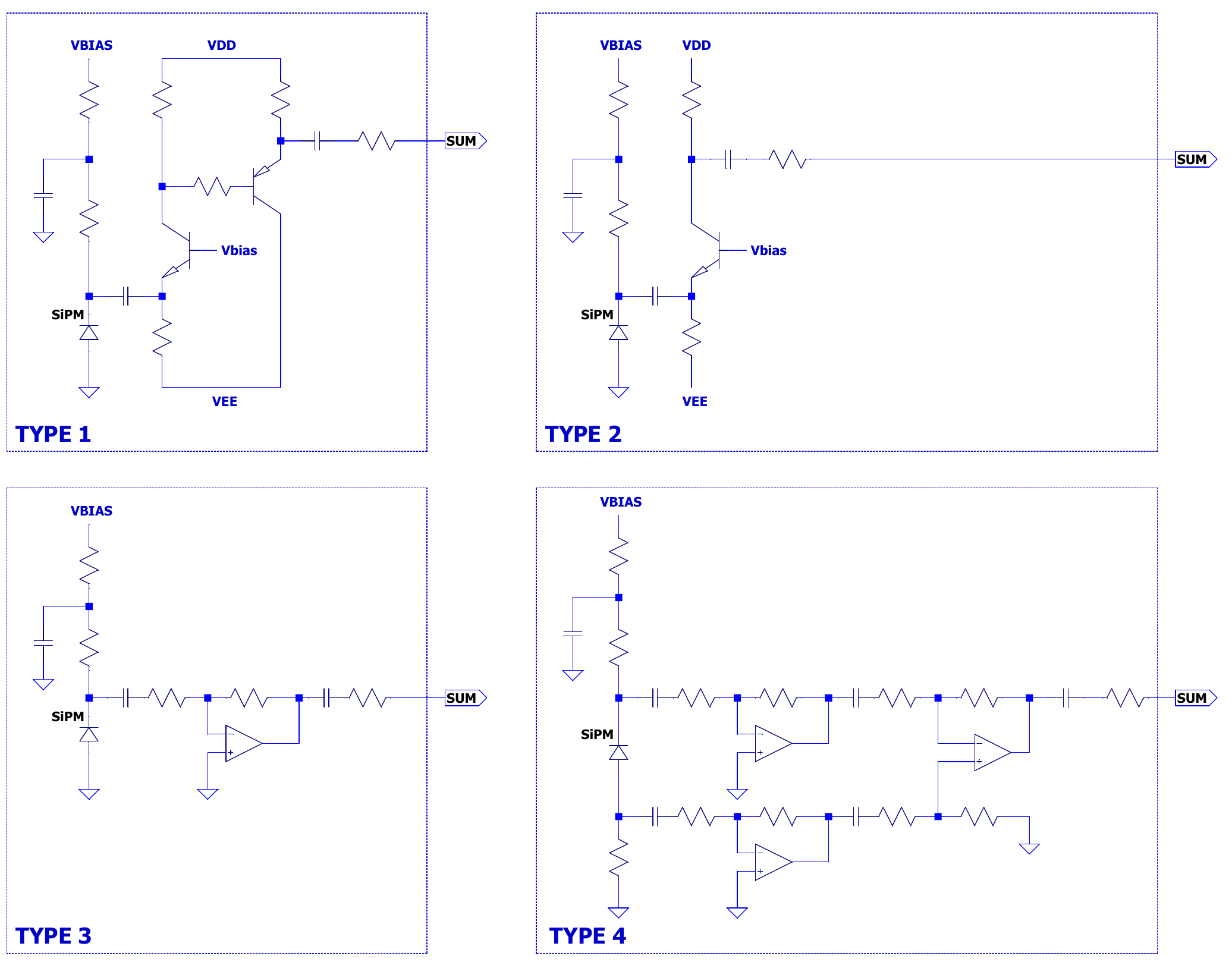}
	\caption{Different tested FEE types. Type 1: buffered common base, Type 2: common base, Type 3: transimpedence, Type 4: differential readout based on transimpedence configuration.}
	\label{fig:circuits}
 \end{center}
\end{figure}

Two basic configurations have been selected for the front-end design: the current feedback and the current conveyor. For each configuration two different readout circuits have been investigated; the basic schematics are shown in Figure \ref{fig:circuits}.

\begin{figure}[ht]
 \begin{center}
	\includegraphics[width=\textwidth]{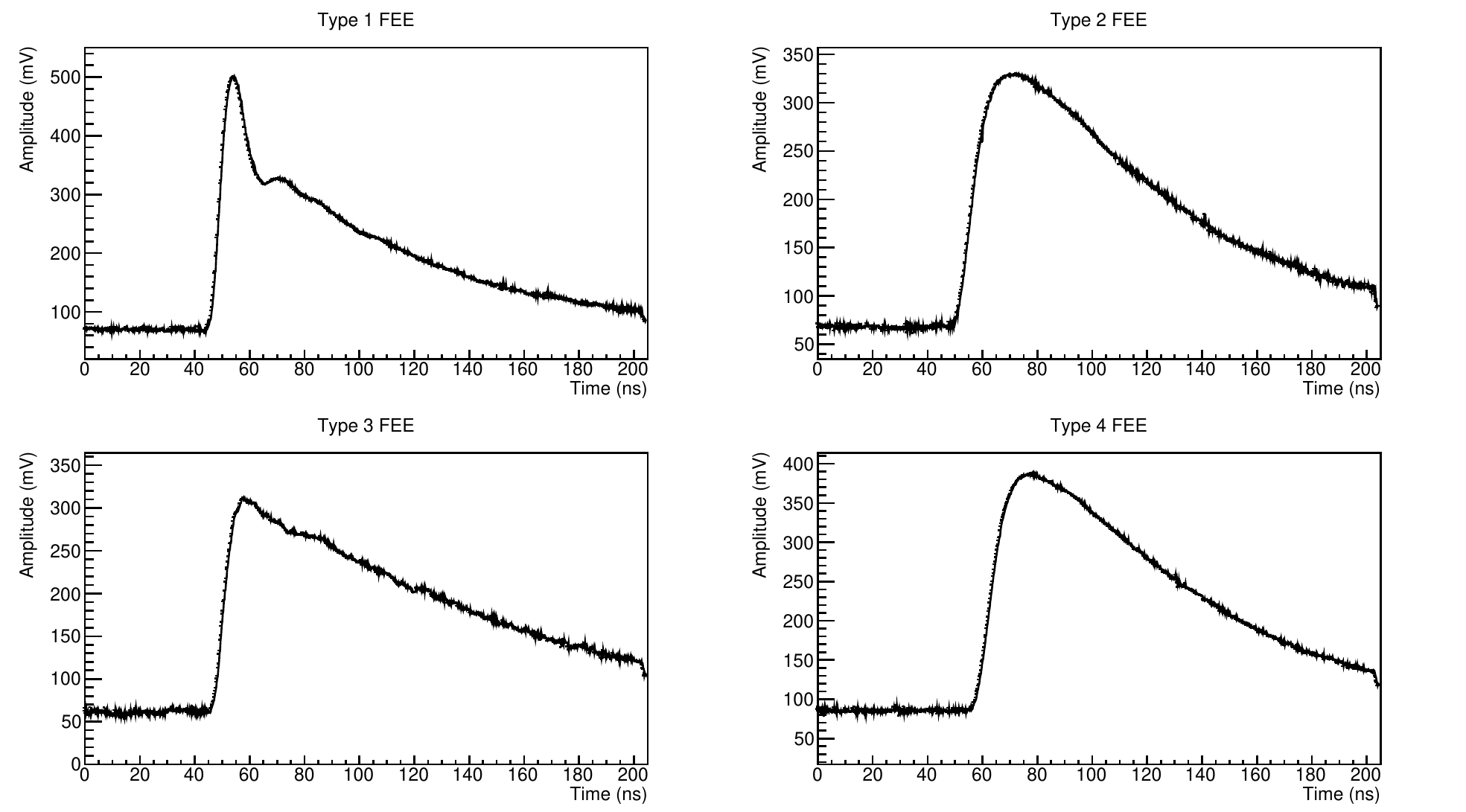}
	\caption{Examples of waveform acquired on tile 4 at 41.5 V bias voltage with the four different FEE.}
	\label{fig:waveforms_cosmic}
 \end{center}
\end{figure}

%
The first two circuits (Type 1 and Type 2) are based on the current conveyor configuration~\footnote{Current Conveyor: $R_c$=collector resistor, $C_{par}$= collector parasitic capacitance} and implemented by means of a NPN BFR92A RF transistor while Type 3 and Type 4 are based on the transimpedance configuration~\footnote{Transimpedance: $R_f$=feedback resistor, $C_{par}$= input parasitic capacitance}. Table~\ref{table:config} shows the main features of the two configurations.  All circuits are AC coupled then allowing baseline fluctuation suppression using low capacitors values; Type 1 circuit includes a local buffer for pole-zero compensation at the summing point while Type 4 configuration helps in suppressing pickup noise (within the amplifier bandwidth). All configurations have been tested in a cosmic ray stand; a detailed description of the  test outcome can be found in section \ref{chapter_cosmics}, while examples of the waveforms collected on the same tile with the four different electronics is shown in Figure \ref{fig:waveforms_cosmic}.

As the four configurations do not show significant differences in the time resolution measurement, the current conveyor one has been selected to carry out the Test Beam, because its resistive input impedance is stable over all the input transistor bandwidth. 
\begin{table}[h!]
\caption{Preamplifier configurations characteristics (a fixed gain has been assumed).}
\centering
\begin{tabular}{c c c} 
\hline\hline
Current Conveyors  & Transimpedance  \\ 
\hline
$Z_{in}$ small (current sensitive) &  $Z_{in}$  $\simeq$ 0 (virtual ground)\\
$Z_{out}$ large (current output) & $Z_{out}$ low (voltage output) \\
Transfer function  $\simeq$ $R_{c}$ & Transfer function  $\simeq$ $R_{f}$ \\
$Z_{in}=1/g_m$ & $Z_{in}=Z_{f}/(G+1)$ \\ 
Bandwidth=1/2$\pi$$R_c$$C_{par}$ & Bandwidth=1/2$\pi$$R_f$$C_{par}$\\
\hline
\end{tabular}
\label{table:config}
\end{table}
%

Finally, to avoid spoiling the signal time information, low impedance connections to the summing point together with a high speed amplifier must be used;  in our design signals have been routed by means of 50 ohm microstrips while the summing point has been implemented  using  the AD8009, a very fast operational amplifier from Analog Devices\footnote{the AD8009 is a high-speed current-feedback amplifier (1 GHz bandwidth) from Analog Devices capable to exploit a 5500 V/$\mu$s slew rate resulting in a sub-ns rise-time if used in low-gain configurations}.

\clearpage

\section{Determination of the intrinsic time jitter of SiPMs and electronics}
\label{sec:time_jitter_sipm}
In order to optimise the overall design of the tiles and corresponding front-end electronics,
it is important to disentangle the individual contributions to the combined time resolution.
The contribution of the SiPM itself combined with the readout electronics has been measured without the scintillator.
A pulse of light, provided by a \texttt{PiL040x}~405~nm laser, was guided onto a single SiPM connected to Type 2 FEE, as in the test beam.
The pulse amplitude was tuned to obtain signals of amplitude comparable to that observed with MIPs.

\begin{figure}[hbt]
  \begin{center}
    \includegraphics[width=0.6\textwidth]{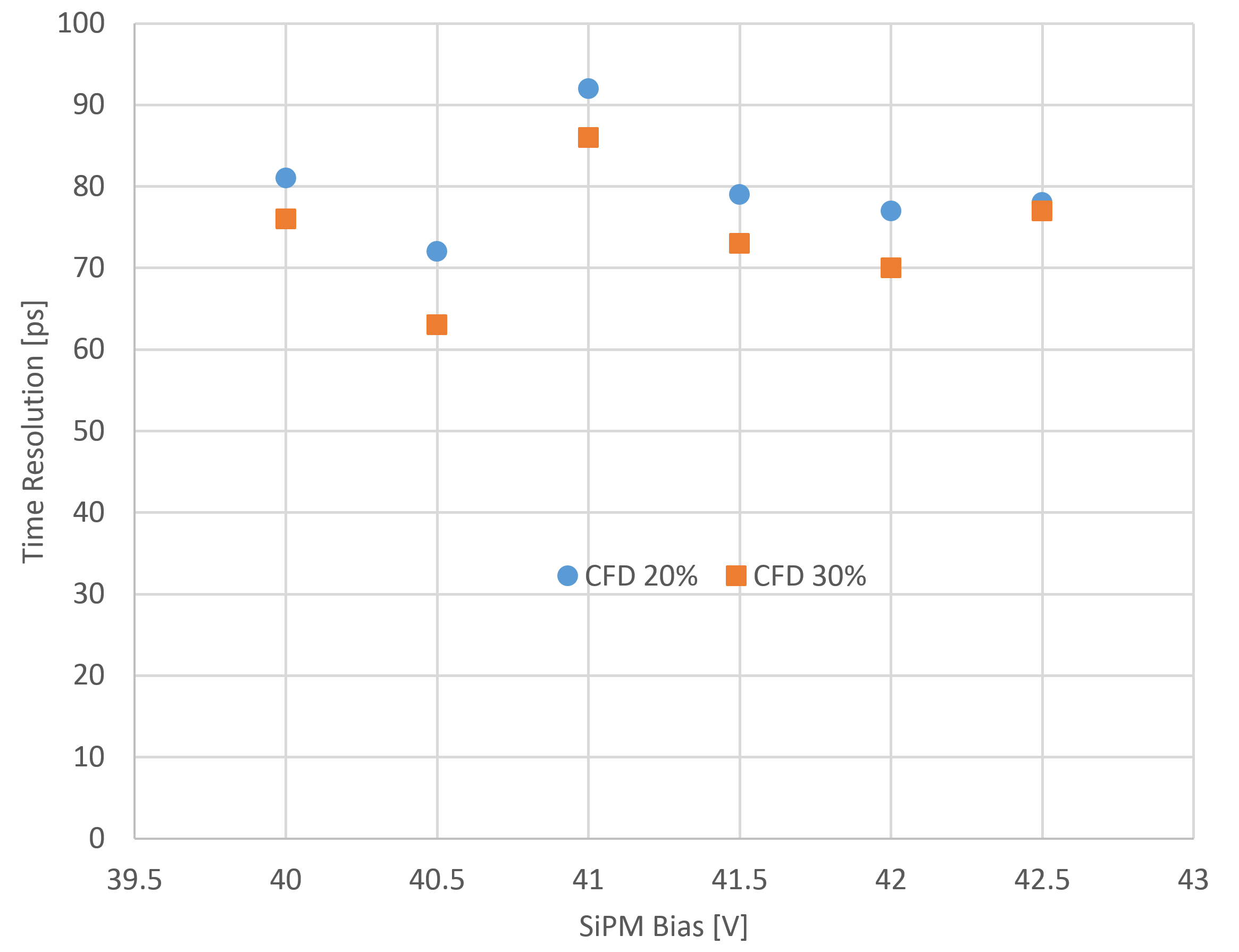}
    \caption{Standard deviation of the difference between laser pulse and SiPM signal, at a given threshold fraction, as function of applied bias voltage.}
\label{fig:laser_sigmat}
\end{center}
\end{figure}

The analog output of the amplifier was acquired together with the sync signal from the laser on a LeCroy WR8000 oscilloscope, operated at 20~GSamples/s.
The oscilloscope was configured to measure the time difference between the two signals, defined as a constant fraction of the SiPM output. Two fractions of the maximum amplitude have been considered,  20\% and 30\%. The standard deviations of the difference, over 1000 laser pulses, are shown in Fig.~\ref{fig:laser_sigmat} as a function of SiPM bias voltage.

Since no obvious dependency was observed, a simple average over all the bias values was taken, yielding a jitter of 74(80)~ps at 30(20)\% constant fraction threshold. Subtracting the full contribution of the laser system (nominally 45~ps), a lower limit for jitter of 66~ps at 20\% threshold was found.

\section{Calibration of the light yield}
\label{sec:light_yield_cal}

In order to express the tiles light yield in number of photo-electrons, the gain of the SiPM and readout electronics has been calibrated by acquiring its signal charge spectrum produced with a low-intensity pulsed led light. The hardware setup used for this measurement consists of a light-tight climate chamber containing a single SiPM connected to the readout electronics.  The light of an LED connected to a pulse generator is directed onto the active surface of the SiPM by an optical fiber, and the amplifier output signal waveforms are acquired by a digital oscilloscope, operated at 10~GS/s. The output charge was measured by integrating the signal waveform within a 200~ns window, similarly to the analysis of test beam data.

\begin{figure}[hbt]
  \begin{center}
    \includegraphics[width=0.8\textwidth]{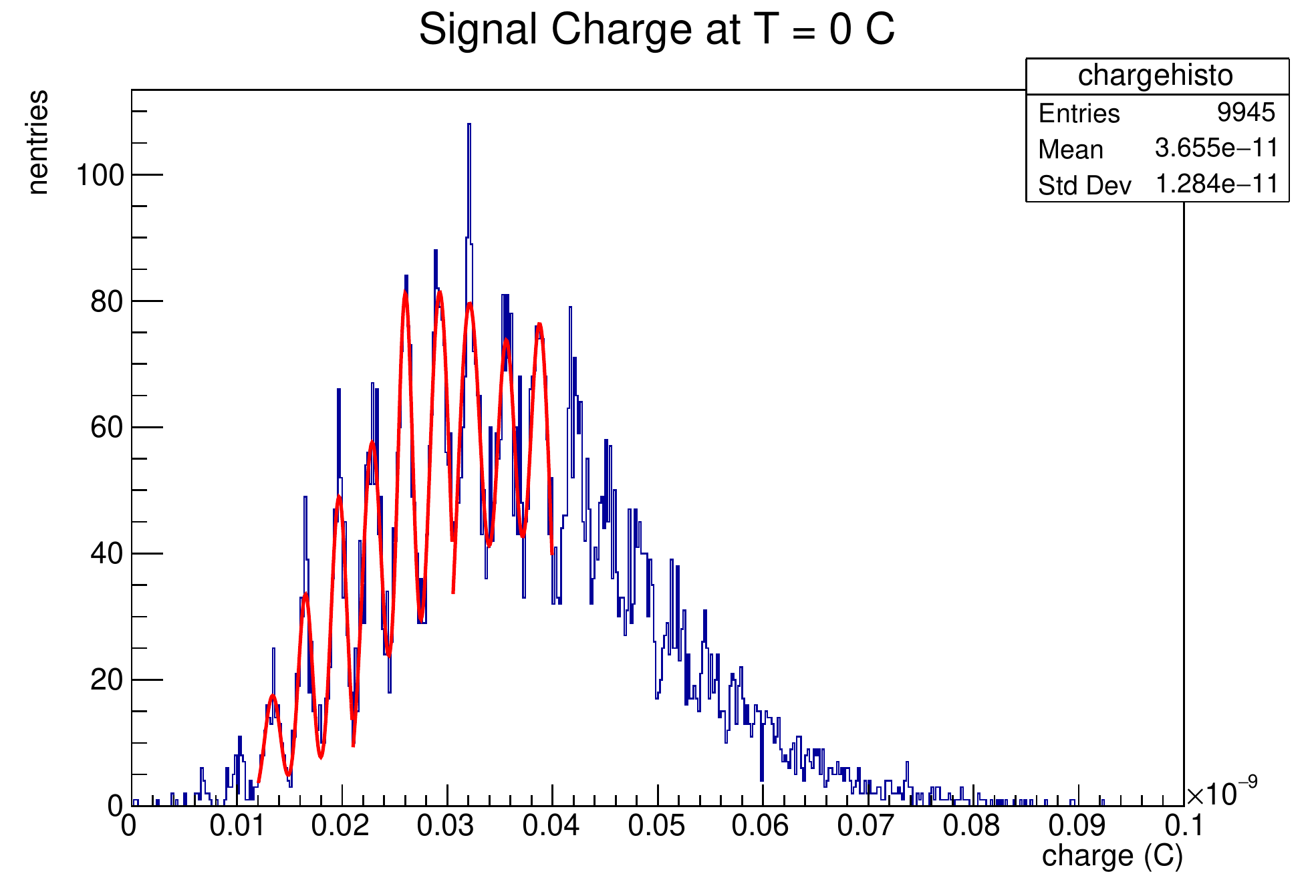}
    \caption{ SiPM  charge spectrum histogram at  0$^o$C. The single photoelectron peaks are fitted with a sum of Gaussian distributions.}
\label{fig:chargespectr}
\end{center}
\end{figure}

To obtain a good separation between the photoelectron peaks of the spectrum, the climate chamber temperature was set to 0$^o$C to reduce the SiPM thermal noise. In order to operate the SiPM at the same gain as at room temperature during the test beam, the bias voltage had to be adjusted to account for the breakdown voltage ($V_{break}$) dependence on temperature, keeping fixed the over-voltage. The $V_{break}$ was determined at 0$^o$C temperature by acquiring the V-I curve with a Keysight B2901A source meter. In a V-log(I) plot, the intersection of the linear fits for the dark current region and the breakdown region provides the sought voltage, yielding a $V_{break}$ of 37.35~V. 
A charge spectrum, shown in Figure \ref{fig:chargespectr}, was acquired and the photoelectron peaks fitted with a sum of multiple Gaussian functions. The average distance between two adjacent peaks corresponds to the signal generated by 1 p.e..  The resulting charge produced by a single photoelectron is $3.2 \pm 0.2$ pC.

\clearpage
\section{Test beam measurements}

\subsection{Test beam setup}
\label{sec:test_beam_setup}

A test beam campaign has been performed at the Beam Test Facility (BTF) of Laboratori Nazionali di Frascati. The BTF is a beam transfer line designed for the optimised, stochastic production of single electrons/positrons for detector calibration purposes. 
Electron and positron beams are created in the energy range of 25-500~MeV with an energy spread of 1\% and a repetition rate varying between 10 and 40~Hz. The typical spot size can vary within (1-25)~mm in $y$ and (1-55)~mm in $x$ and the divergence within 1-2~mrad.
Beam characteristics (spot size, divergence, momentum resolution) are strongly dependent by multiplicity (number of particles/bunch) and energy requested.
For the test beam the chosen configuration was an electron beam of 450~MeV energy with an average particle multiplicity per bunch of 1.8  and a spot size of $\sigma_x = 1.4$~mm and $\sigma_y = 0.8$~mm, as evaluated at the beam pipe exit \cite{btfprofile}.
The beam spot size on the tiles under test, due to the distance from the beam pipe exit, was of few mm$^2$. 

\begin{figure}[hbt]
  \begin{center}
  \includegraphics[width=0.68\textwidth,  height=10cm]{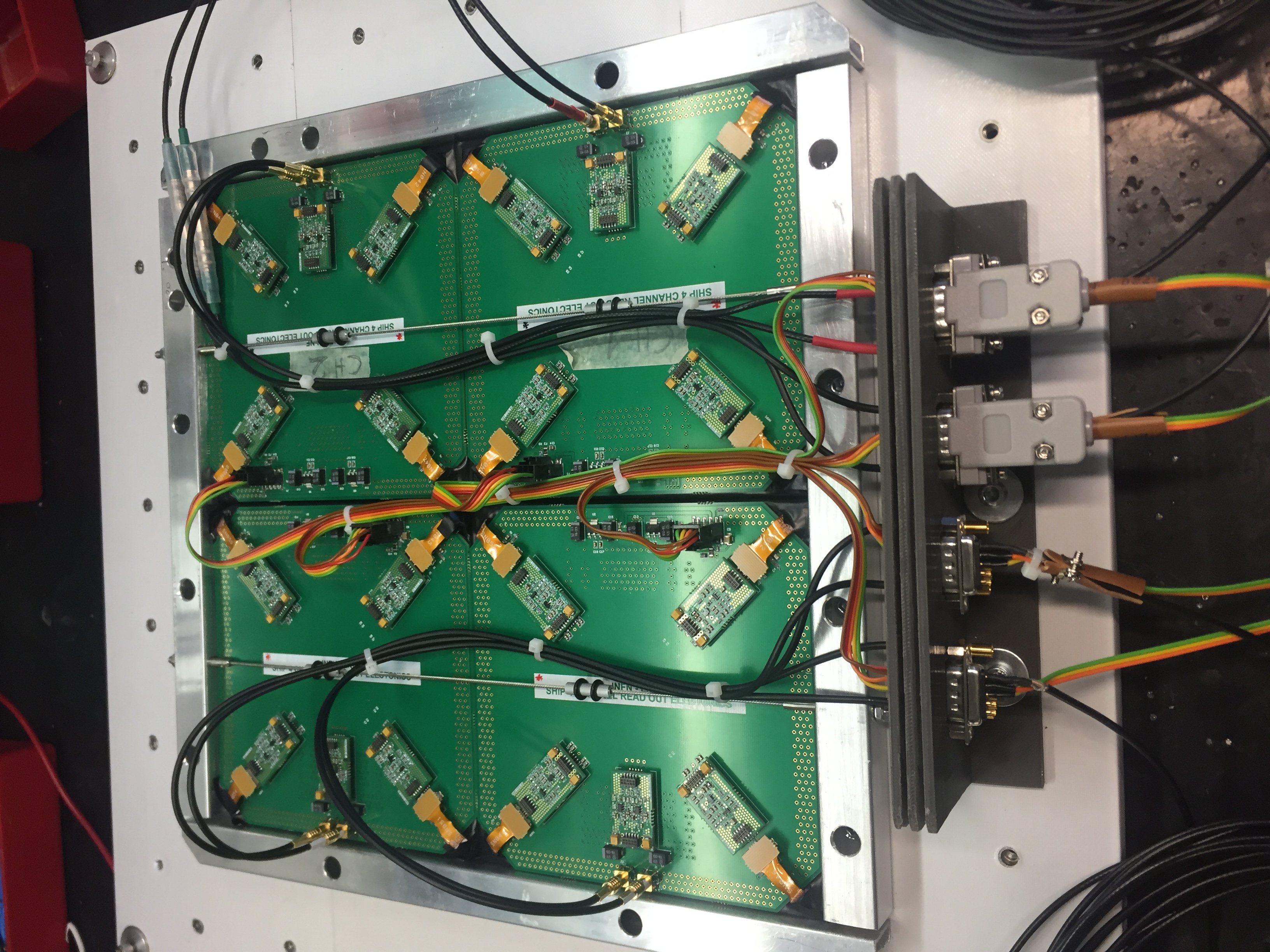}
    \caption{Picture of the minimodule exposed in the test beam.}
\label{fig:minimodule}
\end{center}
\end{figure}

\begin{figure}[hbt]
  \begin{center}
      \includegraphics[width=0.48\textwidth,  height=6cm]{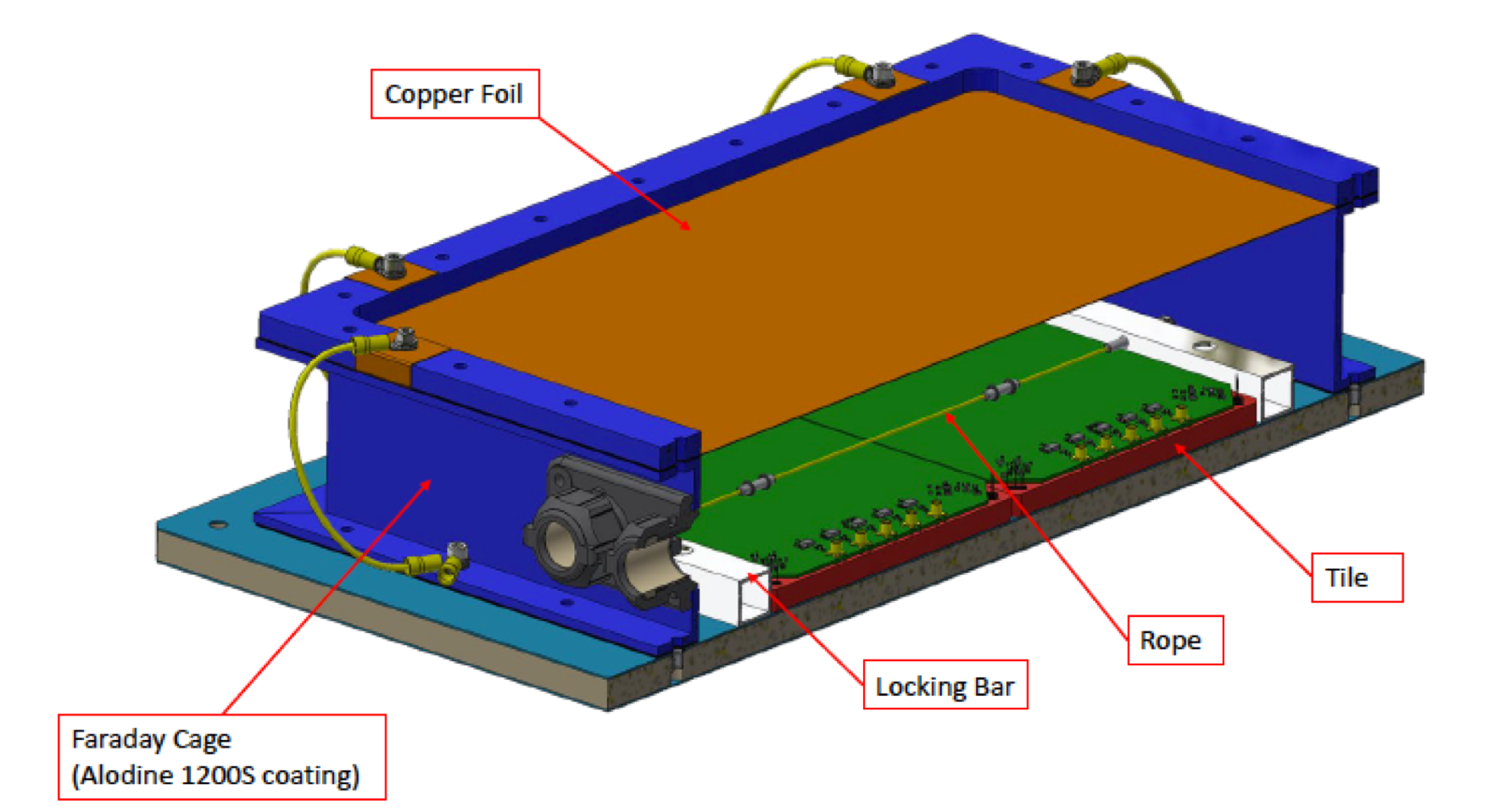}
  \includegraphics[width=0.48\textwidth,  height=6cm]{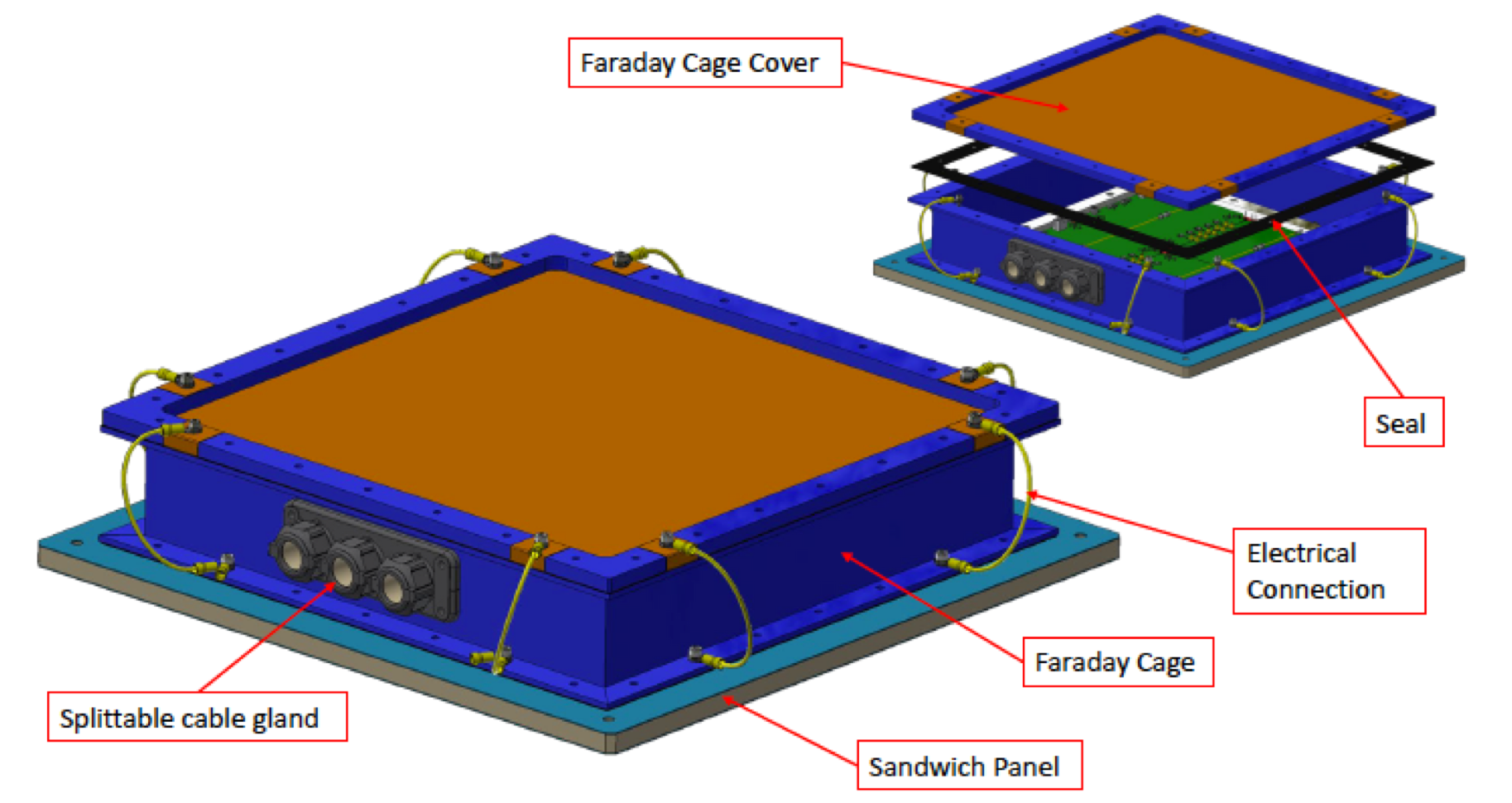}
    \caption{Layout of the light-tight box hosting the four tiles tested at the test beam. See the text for the description.}
\label{fig:box}
\end{center}
\end{figure}

\vskip 2mm
During the test beam campaign the four tiles equipped with Type 2 FEE were hosted into a light-tight box that acted also as a Faraday cage. The system of four-tiles + box, called {\it mini-module}, is shown in Figure~\ref{fig:minimodule}.  
The layout of the box is shown in Figure~\ref{fig:box}. The coverage of the box is provided by 0.5~mm thick copper foil, to minimize the probability of production of electromagnetic showers by the electrons during the test beam. The tiles are kept fixed to the frame by stainless steel wires connecting the two sides of the frame. A patch panel with four 9 way D-sub connectors allows the bias voltages for the SiPMs (one voltage per tile) and the low-voltages for the FEE to be brought inside the box, and the output signals to be sent to an external digitizer.
During the test beam, the SiPMs were powered at a bias voltage of $V_{bias}$=41.5 V at a temperature of about 26$^o$C.
Signals were digitized by a 12-bit 5-GS/s waveform digitizer VME module CAEN V1742.

\begin{figure}[hbt]
  \begin{center}
    \includegraphics[width=0.48\textwidth, height=9cm]{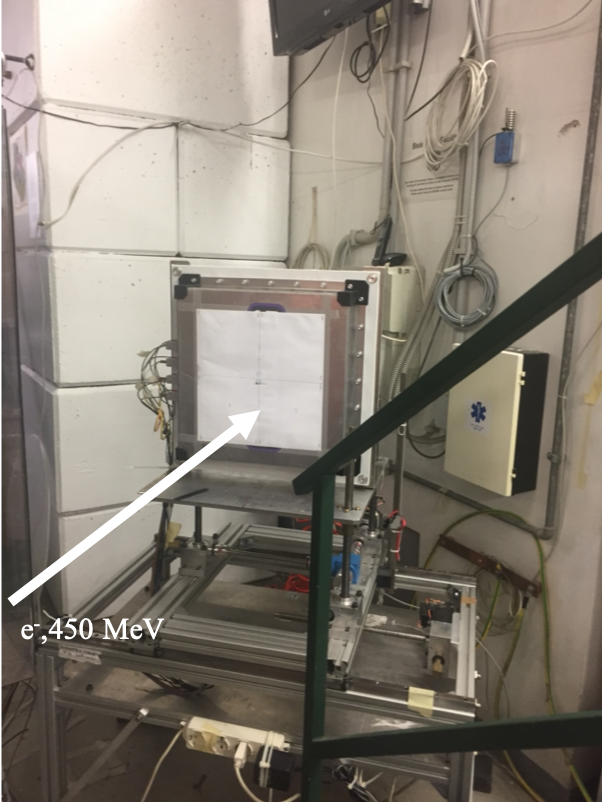}
    \includegraphics[width=0.48\textwidth, height=9cm]{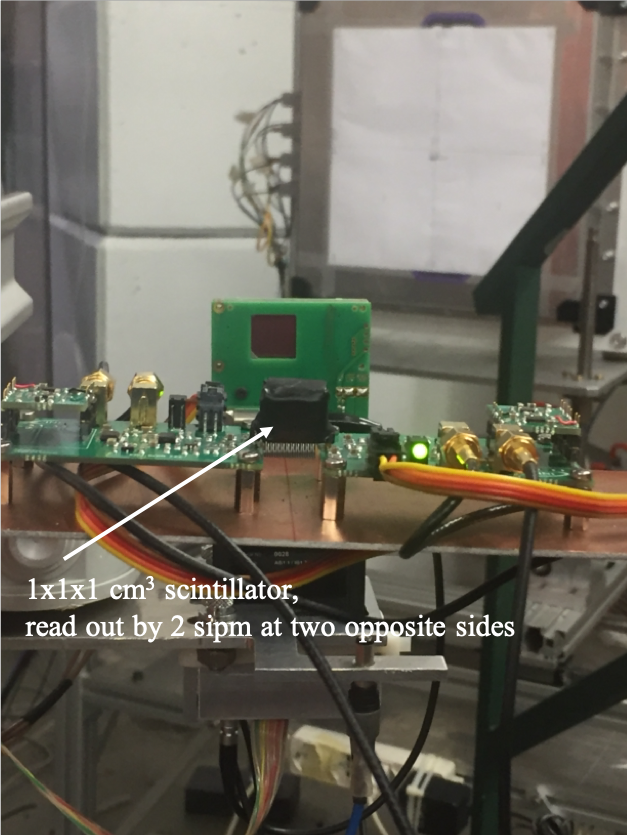}
 \caption{Test beam setup with the four tiles hosted in the box and placed on the moving frame in front of the beam exit (left). Picture of the small scintillating cube used as trigger and time reference (right).}
\label{fig:box_on_beam}
\end{center}
\end{figure}

\vskip 2mm
The mini-module has been exposed to the beam through a rigid frame movable using a remote control with millimetric resolution in order to study the time response and light yield as a function of the beam impinging point. The box was fixed to the moving frame allowing a rigid $x,y$ translation of the position of the tiles with respect to the beam.
The test beam setup is shown in Figure~\ref{fig:box_on_beam} (left).

 The time-reference was provided by a small cube made of EJ200 scintillator with dimensions of $(1 \times 1 \times 1)$~cm$^3$  read out by two $(3 \times 3)$ mm$^2$ Hamamatsu SiPMs S13360-3050CS \footnote{https://www.hamamatsu.com/eu/en/product/type/S13360-3050CS/index.html}, glued on opposite sides of the cube (see Figure~\ref{fig:box_on_beam}, right). The two SiPMs signals were amplified and then digitized by the same digitizer used for processing the tile signals.
 The logic "and" of the two signals from the small cube and the beam radio-frequency signal provided the trigger to the external CAEN digitizer to start the acquisition.

\subsection{Test beam results}
\label{sec:tbresults}

\begin{figure}[hbt]
  \begin{center}
    \includegraphics[width=0.45\textwidth]{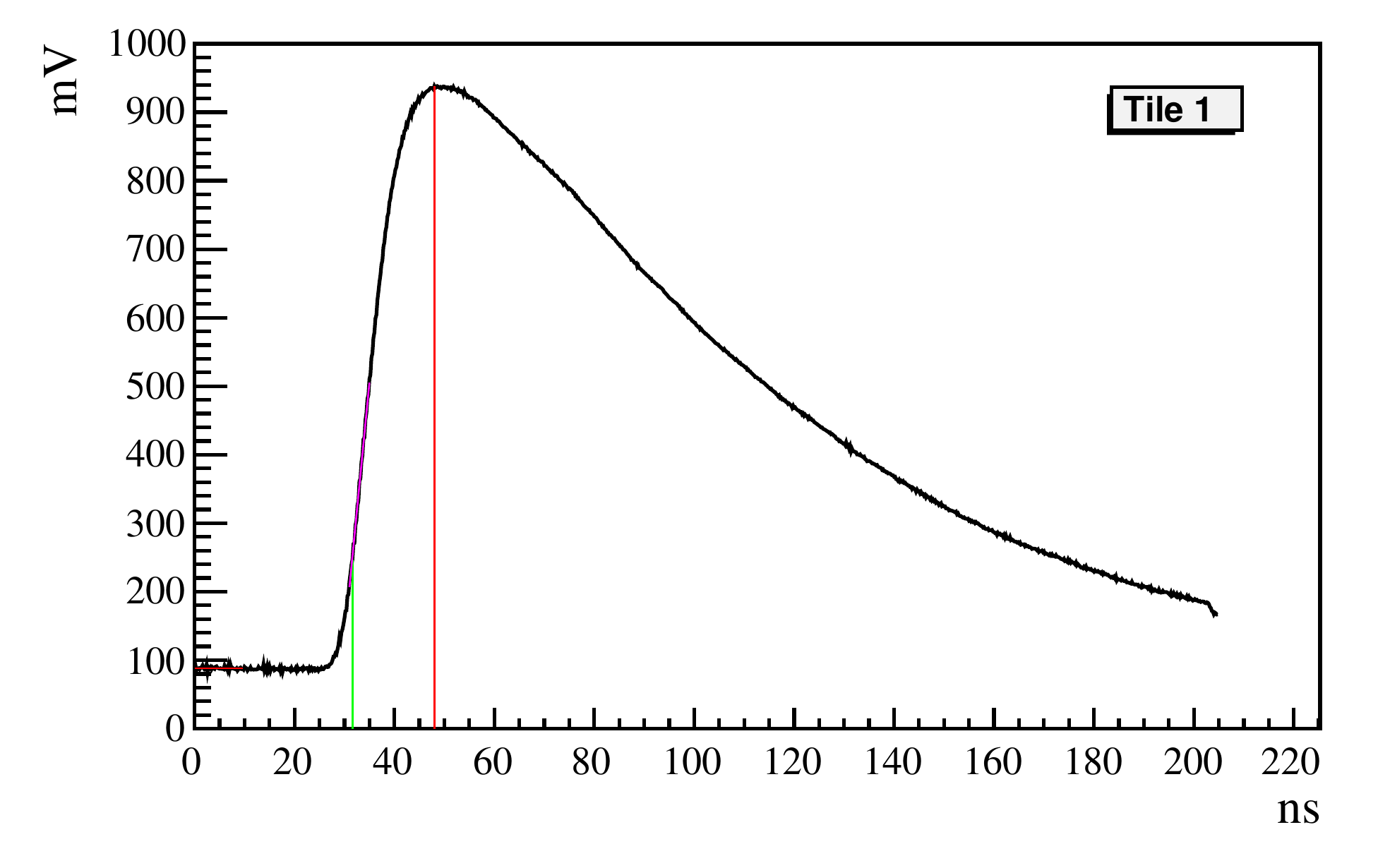}
    \includegraphics[width=0.45\textwidth]{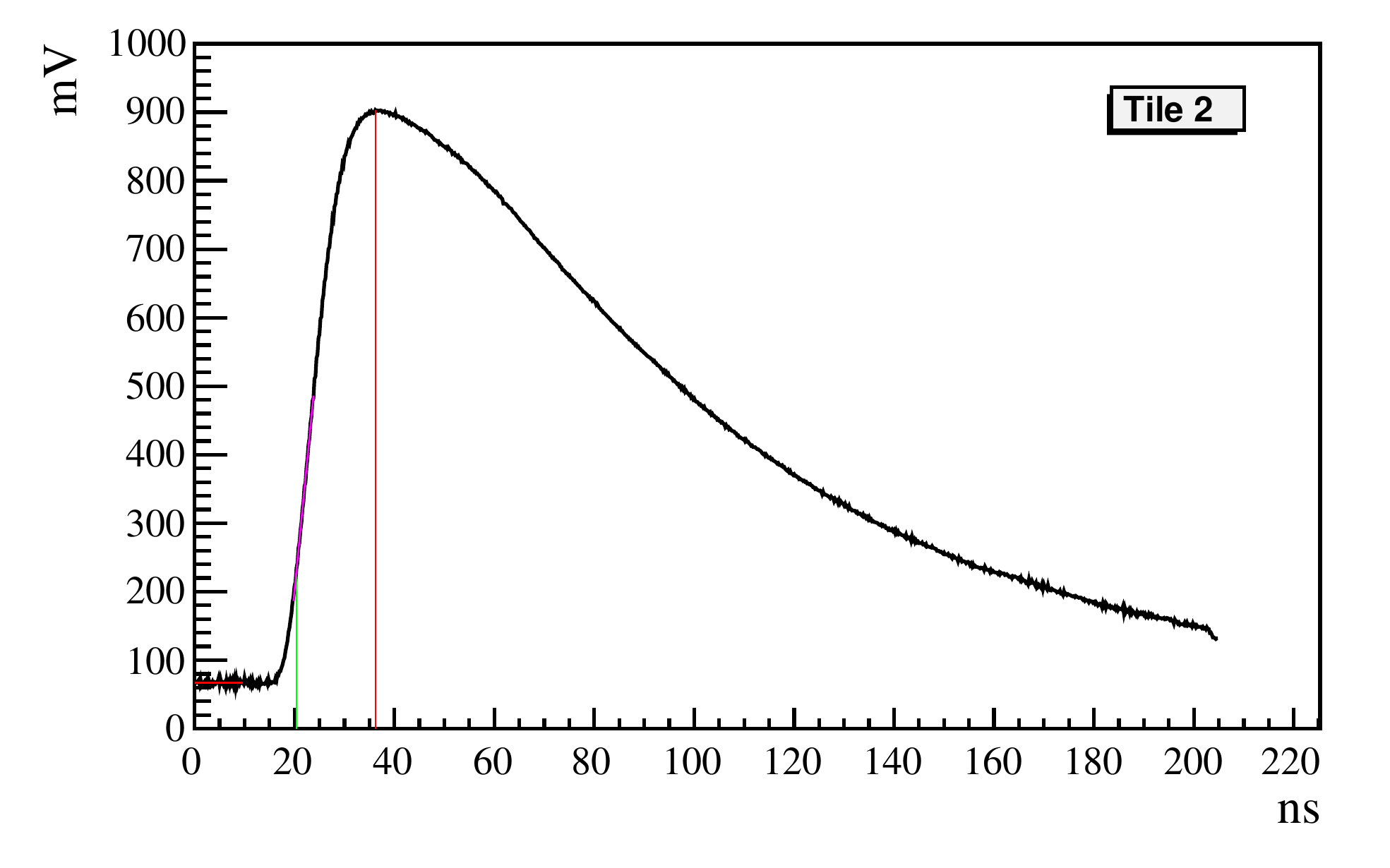}
    \includegraphics[width=0.45\textwidth]{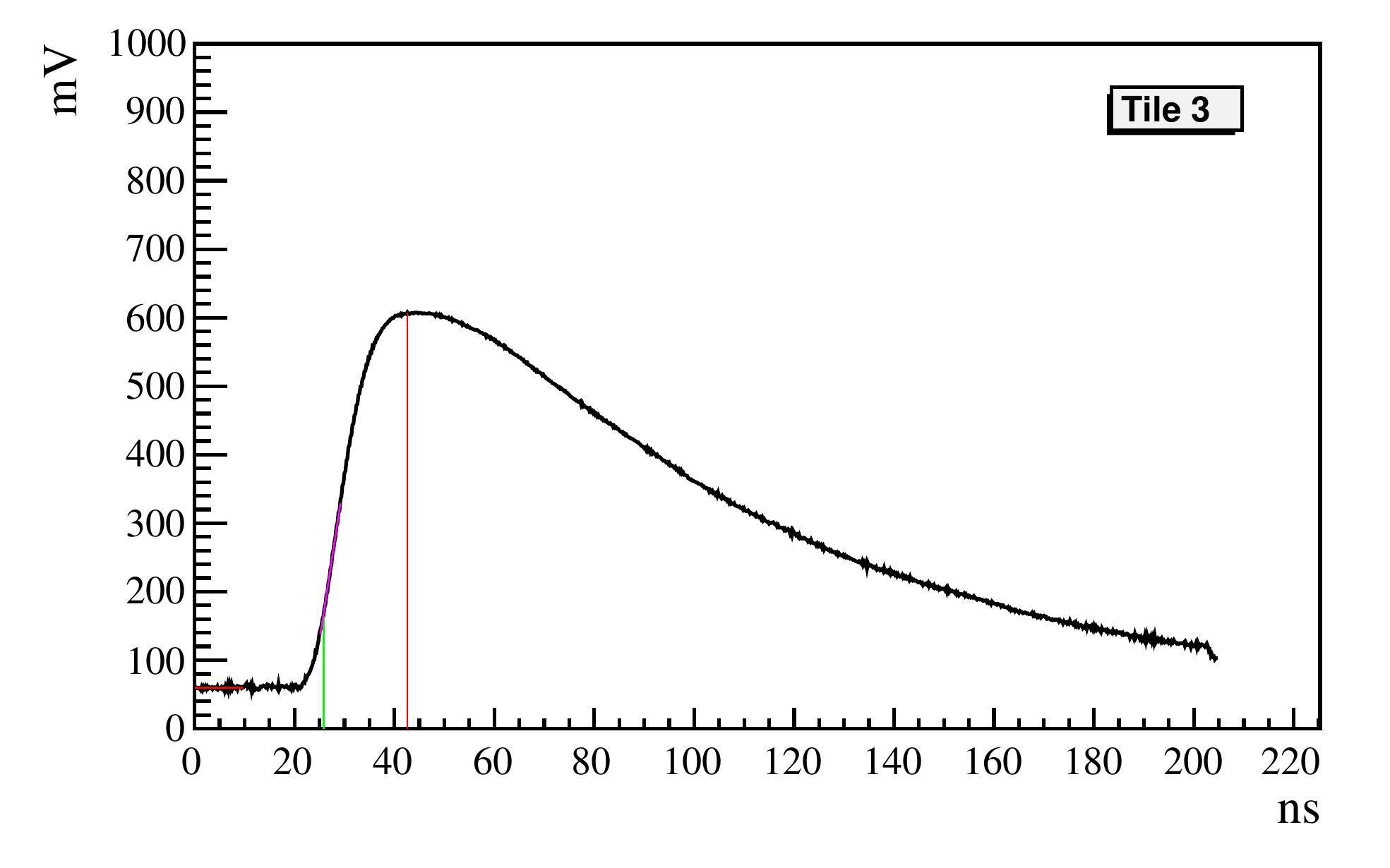} \includegraphics[width=0.45\textwidth]{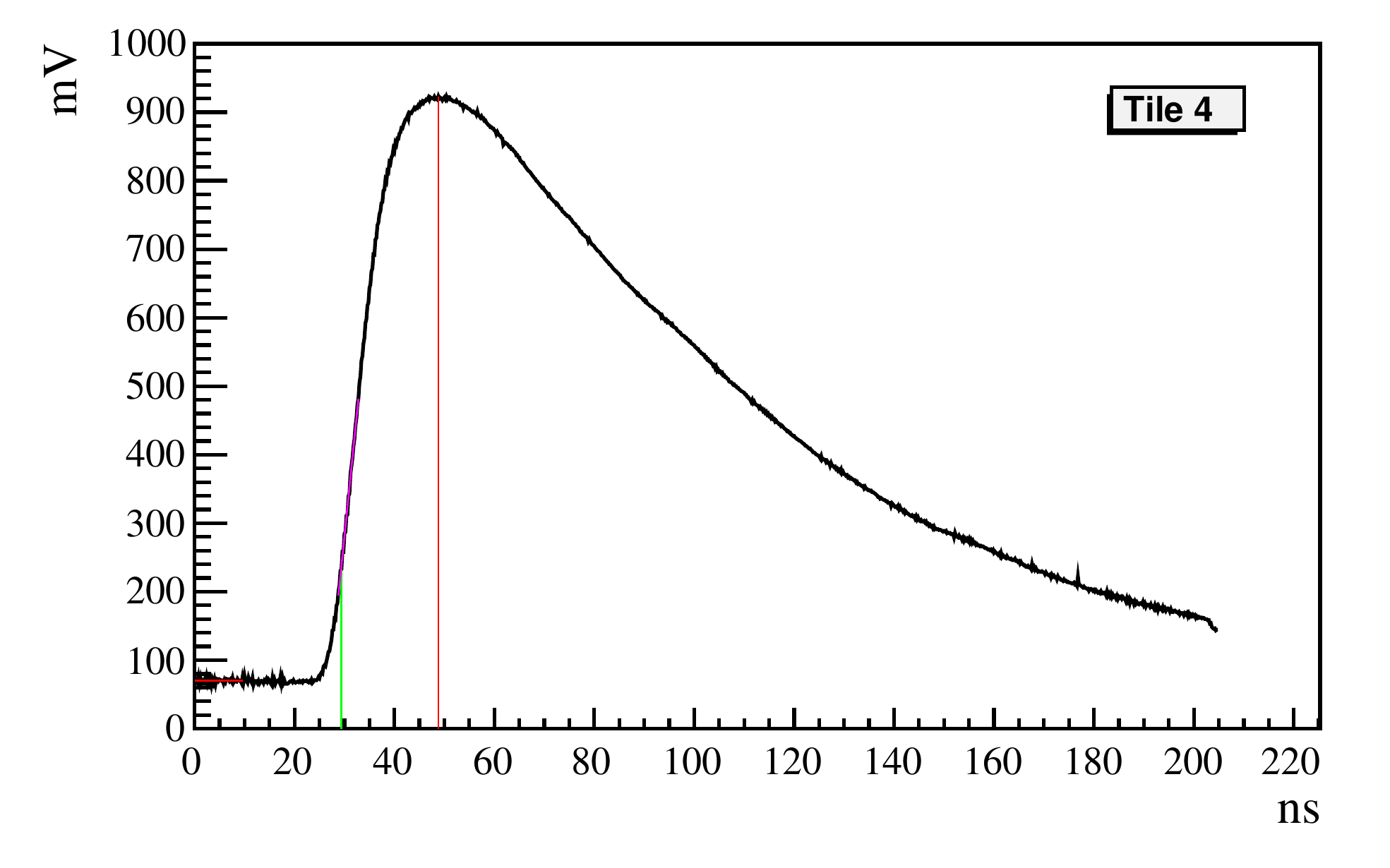}
\caption{Example of signals for the four tiles.}
\label{fig:waveforms}
\end{center}
\end{figure}

Figure~\ref{fig:waveforms} shows an example of signals from the four tiles as recorded by the CAEN digitizer.

\begin{figure}[hbt]
  \begin{center}
    \includegraphics[width=0.45\textwidth]{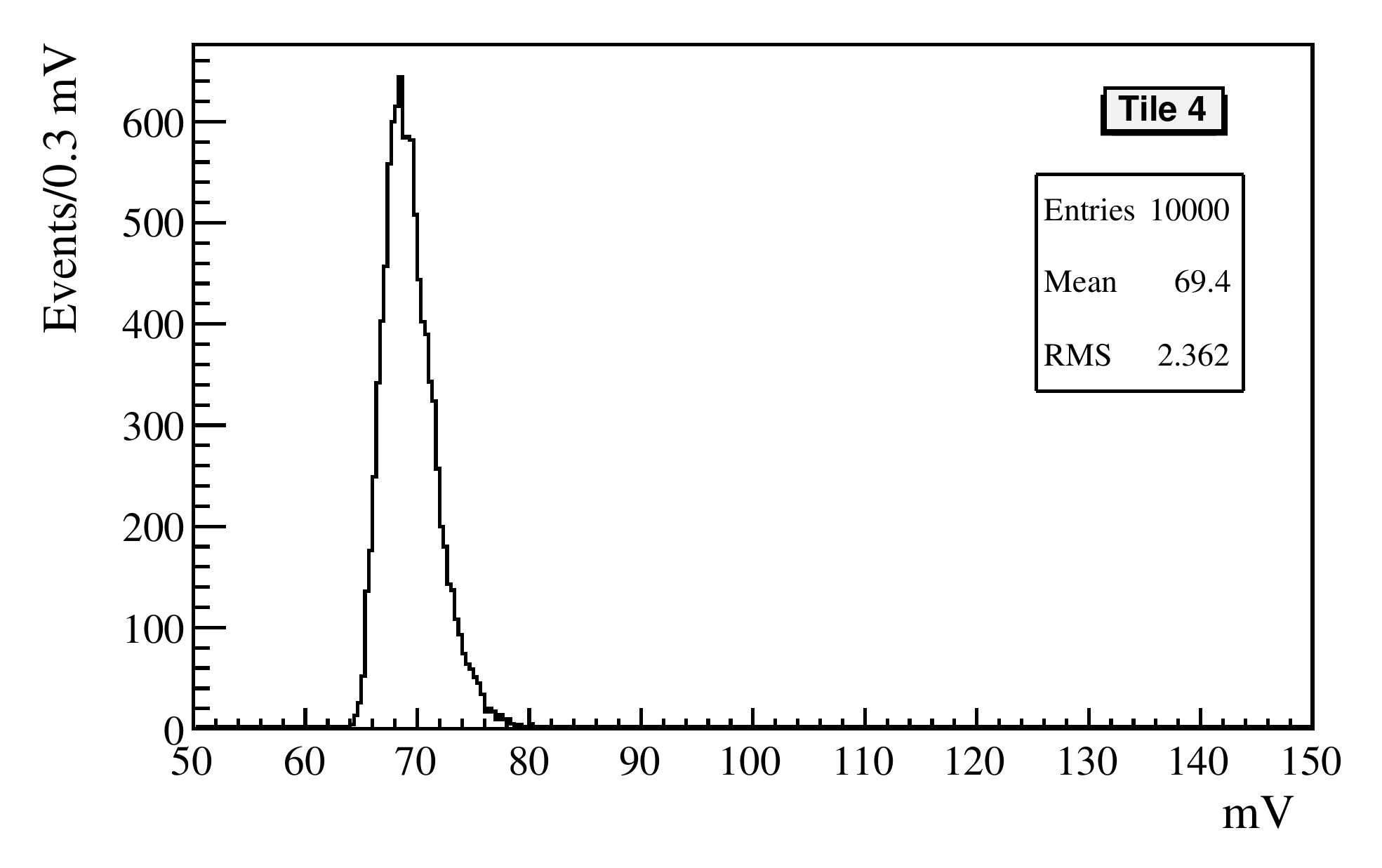}
    \includegraphics[width=0.45\textwidth]{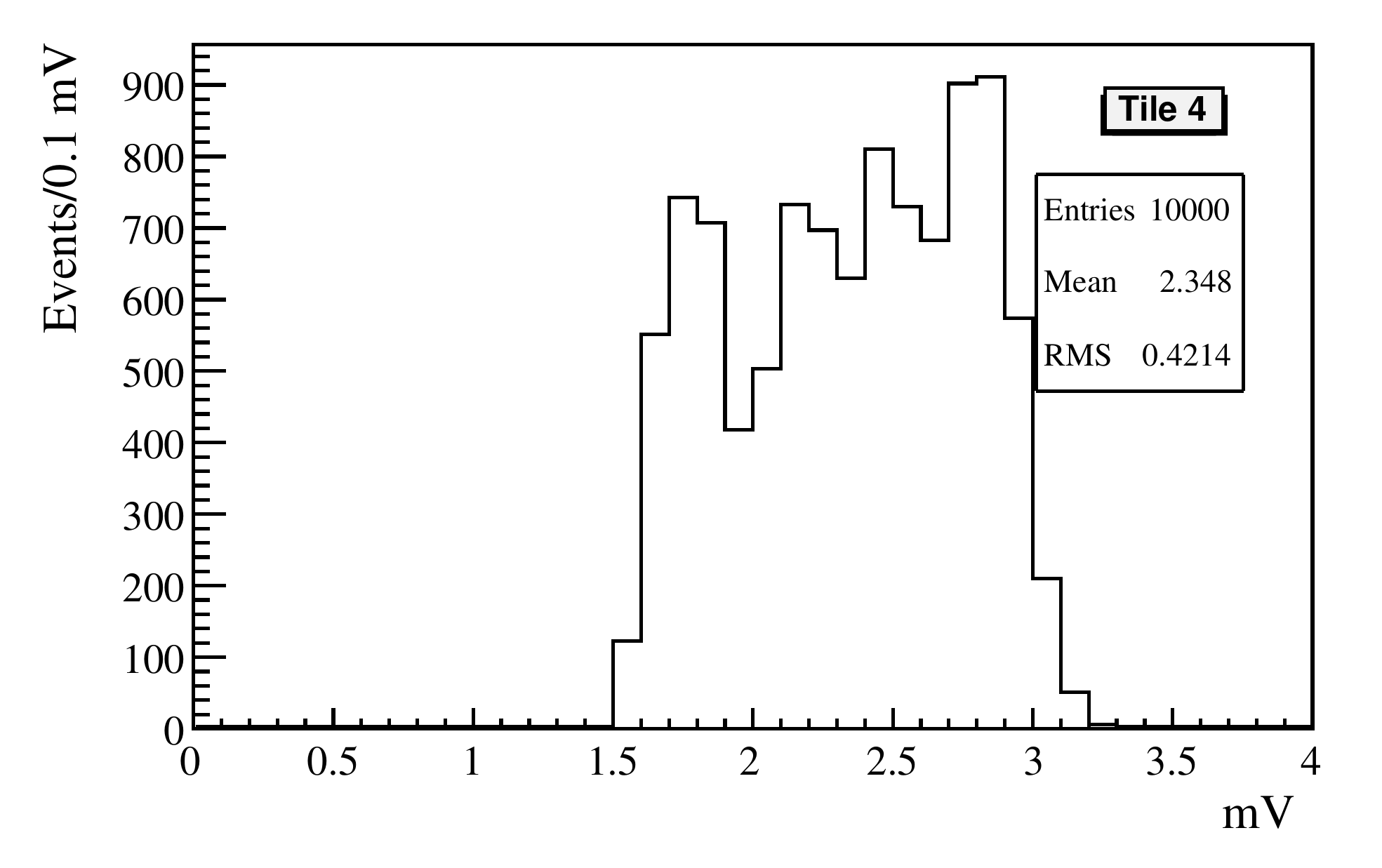}
\caption{Tile 4: Typical distributions of the baseline average value (left) and of the rms of the samples used for its estimation (right).}
\label{fig:baseline_study}
\end{center}
\end{figure}

A linear fit between 2~ns and 10~ns of the waveforms allows the
determination of the mean baseline value, $\mu_{\rm base}$, and the rms of the samplings around it, ${\rm rms}_{\rm base}$.
Examples of the distributions of $\mu_{\rm base}$ and ${\rm rms}_{\rm base}$ are shown in Figure~\ref{fig:baseline_study} for Tile 4. Similar distributions are obtained for the other tiles.
The fluctuation of the $\mu_{\rm base}$ gives an estimate of the low-frequency noise of the system while the average value of the ${\rm rms}_{\rm base}$ is an estimator of the high-frequency one.
The low-frequency noise is eliminated by the event-by-event measurement of the average value of the baseline. The high-frequency noise of the system, instead, affects directly the spread in the determination of the arrival time of the particle and must be kept very low.  The measured value of $\sim 2$~mV with an uncertainty of $\sim 0.5$~mV matches well the requirements.

\vskip 2mm
The arrival time of a particle with respect to the time reference provided by the trigger is measured as follows. The leading edge of the SiPM signal is fitted between
$T_1$ and $T_2$, where $T_1$ is the time in which the signal reaches the 20\% of the maximum amplitude minus 0.6 ns, and $T_2 = T_1 + 4$~ns. The slope is then extrapolated to the baseline. The intercept between the slope and the average baseline value provided by the baseline linear fit defines the arrival time of the particle. This procedure allows the determination of the arrival times to be almost independent on the amplitude value.

\begin{figure}[hbt]
  \begin{center}
    \includegraphics[width=0.48\textwidth]{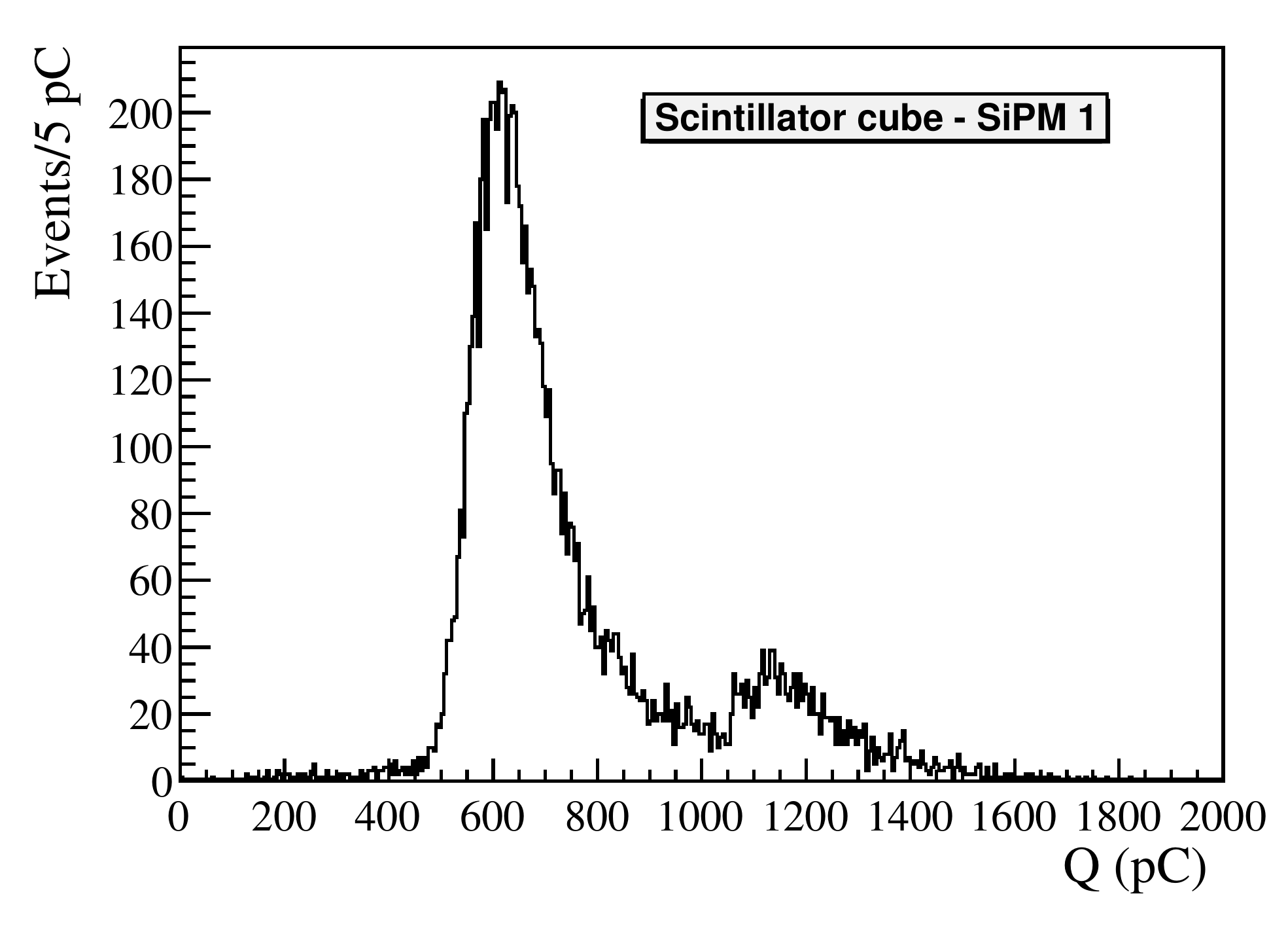}
    \includegraphics[width=0.48\textwidth]{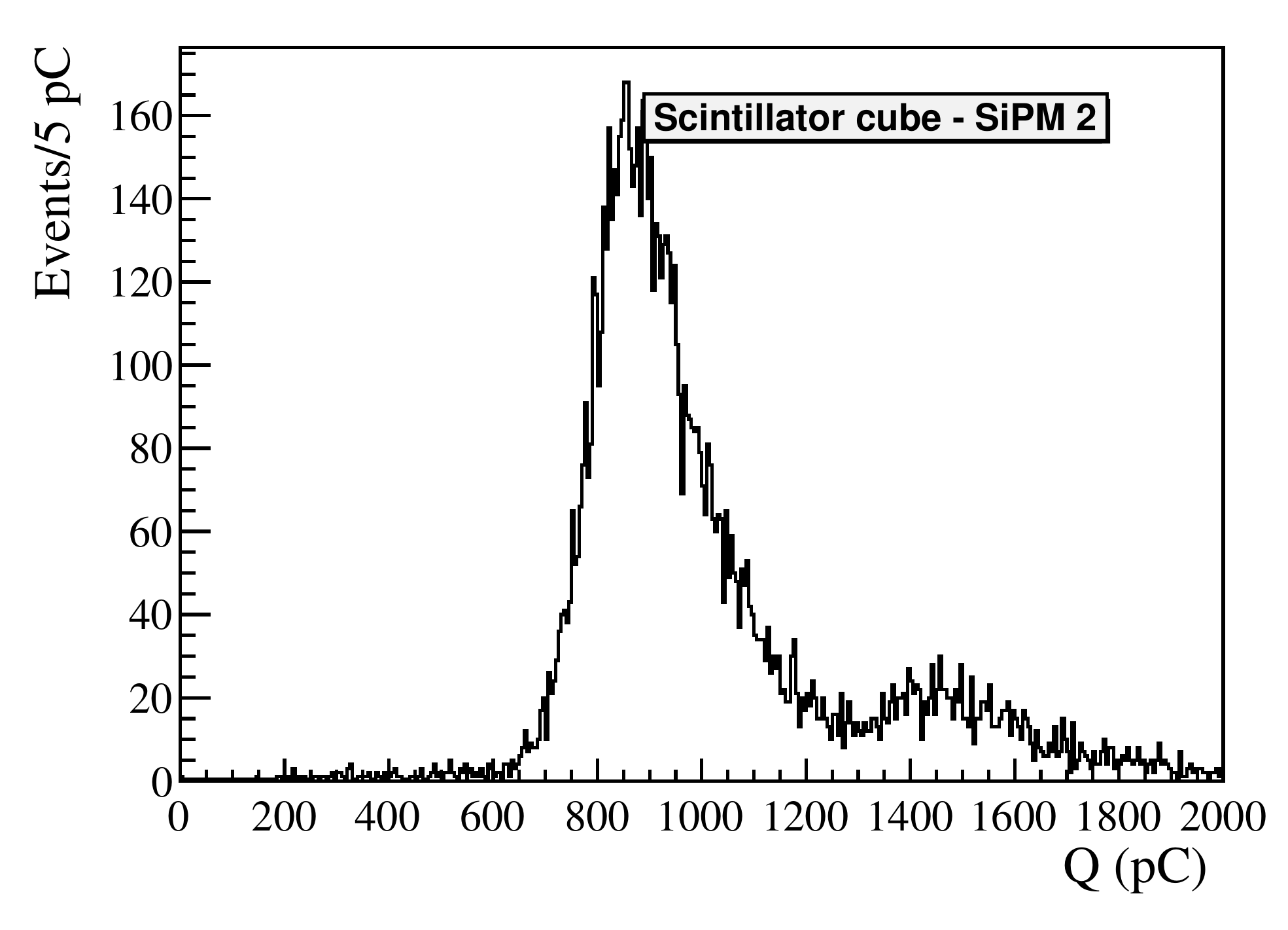}  
\caption{Charge spectra measured by the two SiPMs of the small cube. Single-particle events are defined by requiring a released charge in the range (450-900)~pQ for the first sipm (left) and (600-1300)~pQ for the second (right).}
\label{fig:spectrum_cube}
\end{center}
\end{figure}

\vskip 2mm
The time reference is provided by the average $T_0 = ( T_1 + T_2) /2$ of the two times measured by the two SiPMs of the small cube that provides also the trigger.
In order to get rid of multiple particle events (the average particle multiplicity per spill is 1.8), a selection on the spectrum of the deposited charge measured by the two SiPMs is applied. The two spectra are shown in Figure~\ref{fig:spectrum_cube}. Single-particle events are defined by the  requirement to have a deposited charge in the range (450-900)~pQ for the first SiPM (left plot) and (600-1200)~pQ for the second (right plot).

\begin{figure}[hbt]
  \begin{center}
    \includegraphics[width=0.8\textwidth, height=7cm]{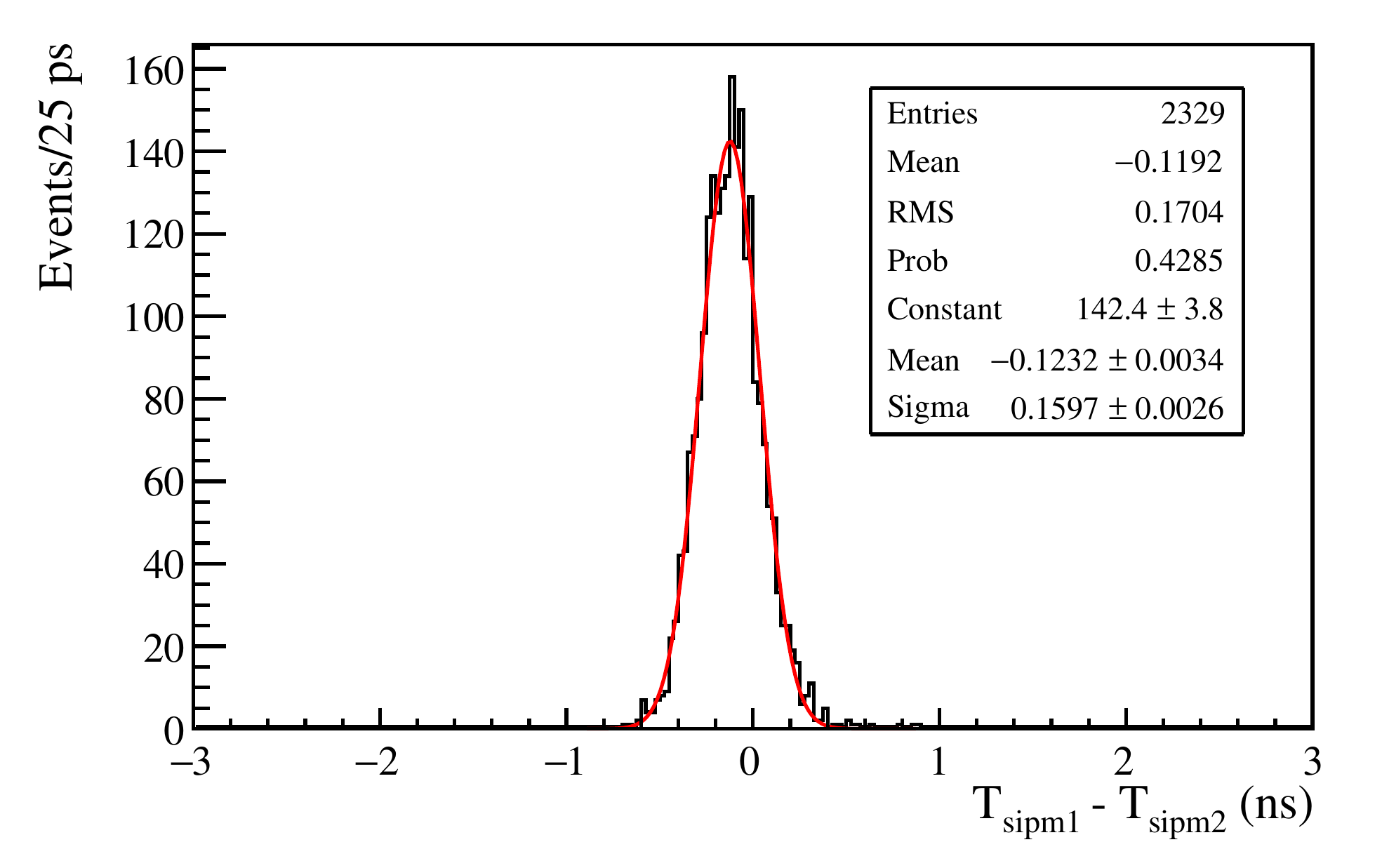}
\caption{Time distribution of the difference in time of the two SiPM signals of the small scintillator cube providing the trigger and the time reference.}
\label{fig:t0_time_res}
\end{center}
\end{figure}

\vskip 2mm
The measurement of the trigger time jitter is extrapolated from the time difference between the signals of the two SiPMs of the cube, $(T_1-T_2)$, whose distribution is shown in Figure~\ref{fig:t0_time_res}. The measurement of $T_1$ and $T_2$ is performed using the same method used for the tiles. 
A value of $\sigma(T_1-T_2)$=160~ps is determined by means of a Gaussian fit.
From this measurement, assuming the two SiPMs and the related FEEs identical, and the time jitter due to the scintillating process uncorrelated between the two, we can evaluate the time resolution of the time reference $T_0 = (T_1 + T_2)/2$, which turns out to be $\sigma(T_0) = \sigma(T_1 - T_2)/2 \sim 80$~ps. 
The time reference resolution must be subtracted in quadrature to the measurement of the time resolution of the tiles.

\begin{figure}[hbt]
  \begin{center}
    \includegraphics[width=0.6\textwidth]{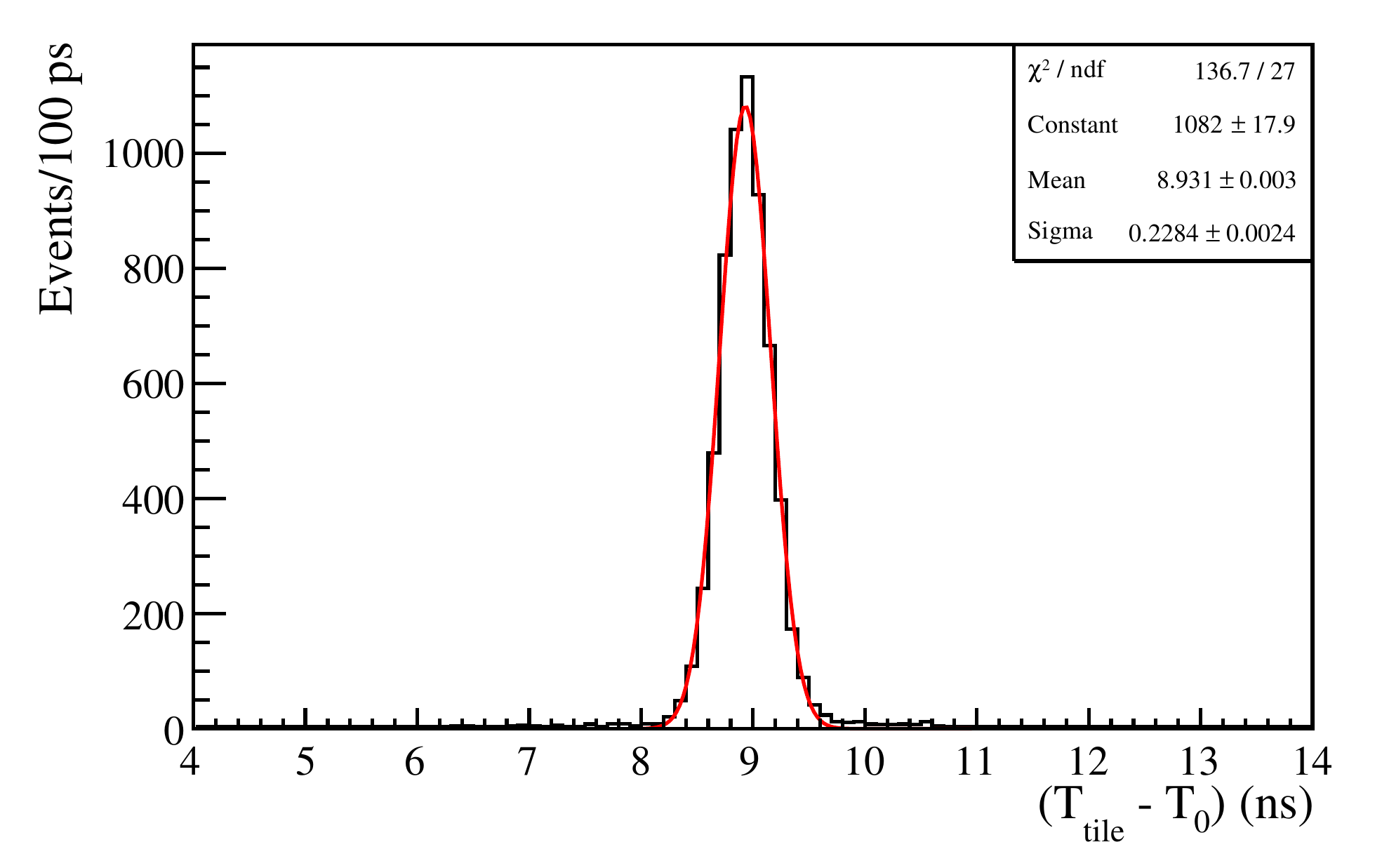}
    \includegraphics[width=0.6\textwidth]{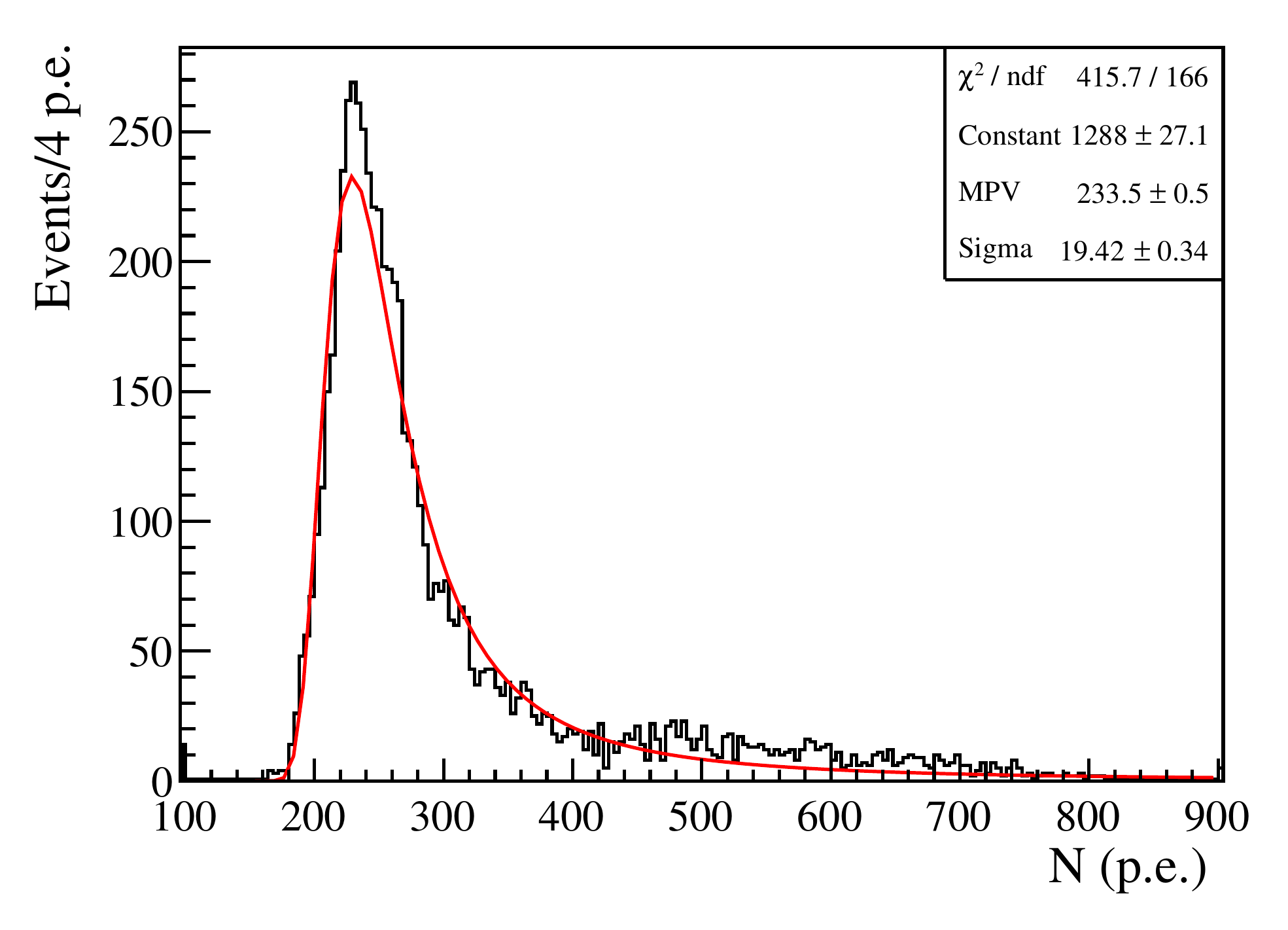}
     \includegraphics[width=0.6\textwidth]{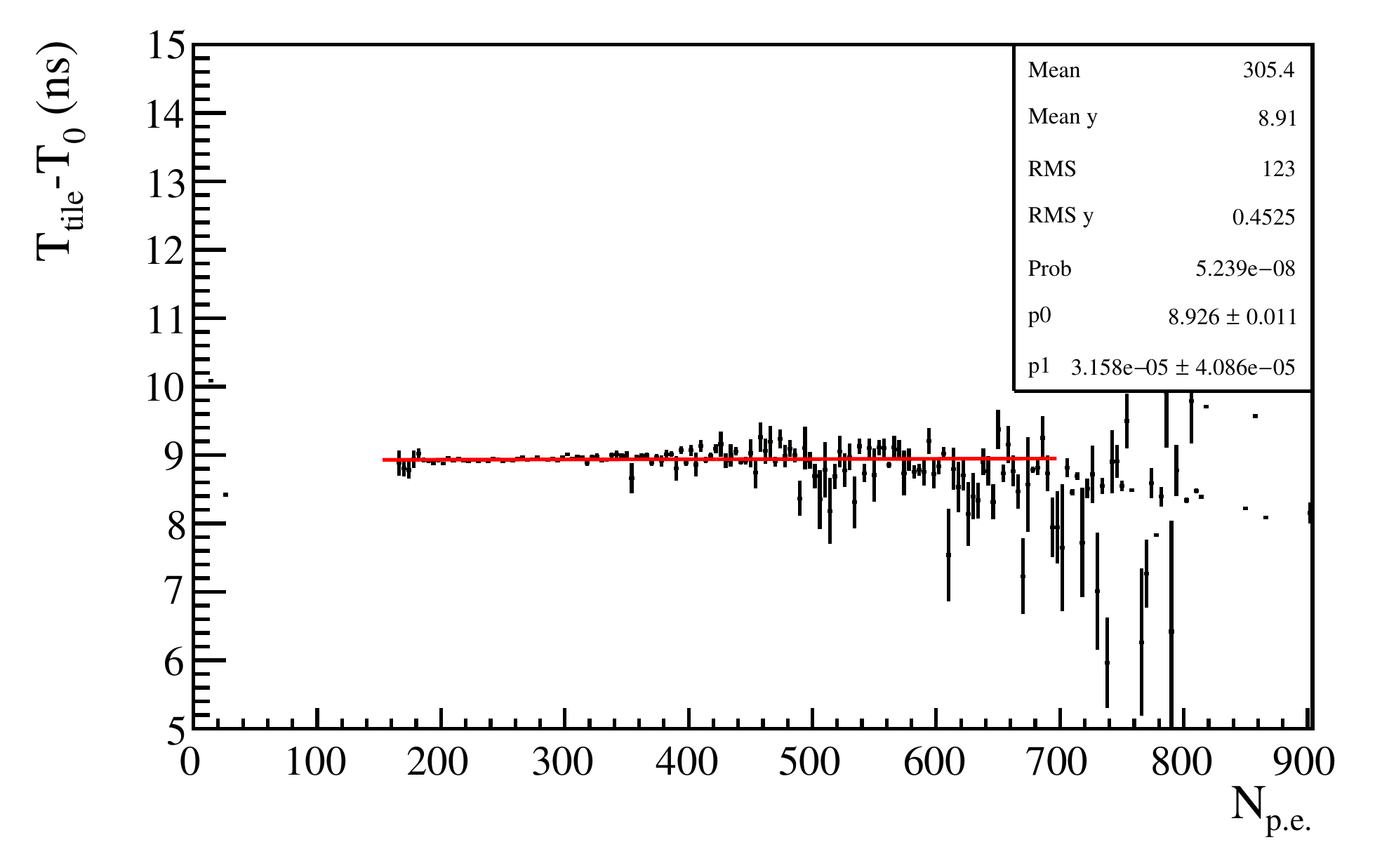}
    \caption{Example of the time arrival distribution for the beam impinging in the tile centre (top), charge spectrum (center), and time arrival average as a function of the deposited charge (bottom ) for the sum of the four SiPM signals on a single tile.}
\label{fig:tile_res}
\end{center}
\end{figure}

\vskip 2mm
An example of arrival time distribution when the beam is at the centre of tile 1 is shown in Figure~\ref{fig:tile_res} (top).
The time resolution is obtained by fitting with a gaussian function the distribution and taking the standard deviation.
After the subtraction of the T0 jitter, the time resolution is estimated to be 215~ps for a beam impinging on a central position.  As expected, the arrival time determination shows a negligible dependence on the deposited charge (see Figure~\ref{fig:tile_res} (bottom)). 
The distribution of the deposited charge is also shown (see Figure~\ref{fig:tile_res} (center)). 

\begin{figure}[hbt]
  \begin{center}
  \includegraphics[width=0.7\textwidth]{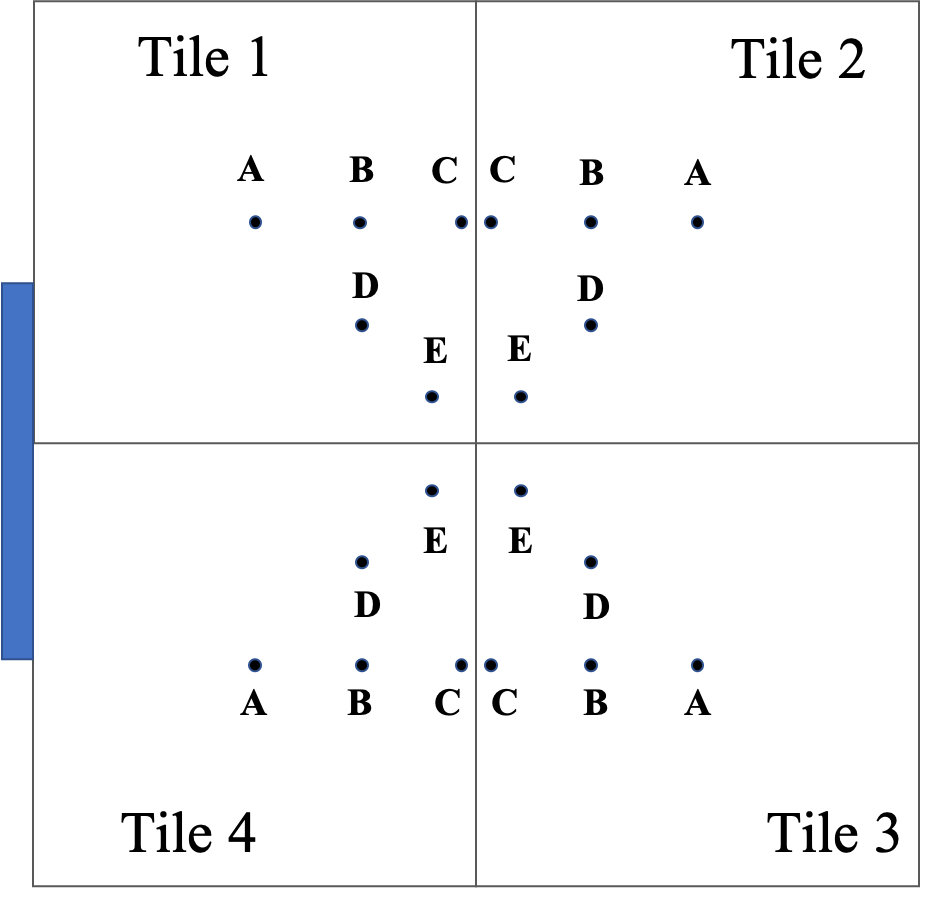}
\caption{Positions of the beam on the mini-module during the scan performed at the BTF test beam.}
\label{fig:scan}
\end{center}
\end{figure}

Figure~\ref{fig:scan} shows the $x-y$ scan over the tiles performed during the test beam. 
Table~\ref{tab:scan_results} shows the results obtained at each point of the scan for arrival time, time resolution and light yield. 
We see that the time resolution is pretty stable (within $\pm$ 10\% maximum variation) among the different tiles and for different beam positions on them, while the average time response shows a drift of $\sim 20\%$ while moving from the tile centre towards the corners.

The impact of this drift on the time resolution has been evaluated using cosmics by illuminating almost uniformly a large fraction of the tile area. 
The results are shown in Section~\ref{chapter_cosmics}.

\begin{table}[htbp]
\caption{Results of the scan over the mini-module. The points used during the scan are shown in Fig.~\ref{fig:scan}. For each of them the positions with respect to the (0,0) coordinate at the mini-module centre, the average time arrival, the resolution of the time arrival, and the light yield are shown. The time resolution is already corrected for the $T_0$ time jitter, while the conversion of the light yield in photo-electrons is performed exploiting the calibration described in section \ref{sec:light_yield_cal}.}
\label{tab:scan_results}
\vspace{.1cm}
\begin{center}
\begin{tabular}{cccccc}
\hline \hline
 Tile  & Label & Position & Average arrival time & arrival time resolution & light yield \\ 
  n.     &  & [cm,cm]     & [ns]       & [ps] &  [N p.e.]  \\ \hline
 1   & A  & [-7.5,7.5] & 8.93$\pm$0.003 & 226$\pm$2  & 230$\pm$20\\
 1   & B  & [-3.5,7.5] & 8.87$\pm$0.003 & 219$\pm$3  & 250$\pm$21 \\
 1   & C  & [-0.5,7.5] & 8.82$\pm$0.003 & 210$\pm$1  & 260$\pm$20\\
 1   & D  & [-3.5,3.5] & 8.572$\pm$0.003 & 243$\pm$3 & 300$\pm$30 \\ 
 1   & E  & [-1.5,1.5] & 7.992$\pm$0.003 & 228$\pm$3 &  520$\pm$80\\   
 \hline
 2   & A  &  [7.5,7.5] & 8.696$\pm$0.003 & 237$\pm$3 & 186$\pm$16 \\
 2   & B  &  [3.5,7.5] & 8.615$\pm$0.003 & 226$\pm$3 & 205$\pm$20 \\
 2   & C  &  [0.5,7.5] & 8.541$\pm$0.003 & 222$\pm$3 & 217$\pm$20 \\
 2   & D  &  [3.5,3.5] & 8.468$\pm$0.003 & 227$\pm$3 & 263$\pm$30 \\ 
 2   & E  &  [1.5,1.5] & 8.542$\pm$0.003 & 223$\pm$3 & 640$\pm$110 \\ \hline
 3   & A  &  [7.5,-7.5] & 9.188$\pm$0.003 &  229$\pm$2 & 207$\pm$18\\
 3   & B  &  [3.5,-7.5] &  8.991$\pm$0.007 &  234$\pm$7 & 230$\pm$22 \\
 3   & C  &  [0.5,-7.5] &  8.943$\pm$0.007 &  239$\pm$6 & 231$\pm$18 \\ 
 3   & D  &  [3.5,-3.5] &  8.797$\pm$0.003 &  237$\pm$3 & 358$\pm$44\\
 3   & E  &  [1.5,-1.5] &  7.87$\pm$ 0.003 &  227$\pm$3 & $680 \pm 200$ \\ 
 \hline
 4   & A  & [-7.5,-7.5] &  10.231$\pm$0.003 & 248$\pm$2 & 138$\pm$13 \\
 4   & B  & [-3.5,-7.5] &  10.154$\pm$0.003 & 247$\pm$2 & 143$\pm$14 \\
 4   & C  & [-0.5,-7.5] &  10.172$\pm$0.003 & 245$\pm$3 & 141$\pm$13 \\
 4   & D  & [-3.5,-3.5] &  9.927$\pm$0.003 & 230$\pm$3 & $183 \pm 20$ \\
 4   & E  & [-1.5,-1.5] &  9.898$\pm$0.003 & 207$\pm$3  & $397 \pm 78$\\

\hline \hline
\end{tabular}
\end{center}
\end{table}

\clearpage
\section{Cosmic rays measurements}
\label{chapter_cosmics}

A cosmic ray test stand has been instrumented at Laboratori Nazionali di Frascati with the threefold aim to:
\begin{itemize}
    \item [i)] measure the time resolution when a large fraction of the tile area is uniformly illuminated;
    \item[ii)] evaluate the tile efficiency for minimum ionizing particles;
    \item[iii)] test the four different FEEs.
\end{itemize}

\begin{figure}[hbt]
  \begin{center}
    \includegraphics[width=0.5\textwidth]{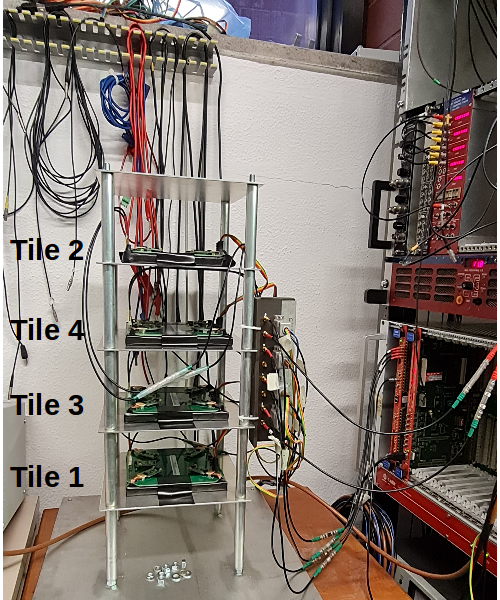}
\caption{Cosmic ray set-up with the four tiles piled up. The tiles are labelled as in table \ref{tab:tiles}.}
\label{fig:cosmic_setup}
\end{center}
\end{figure}

The four tiles have been piled up at a distance of 10 cm one from the other, as shown in Figure~\ref{fig:cosmic_setup}.

The two EJ200 tiles have been positioned at the edges, at a mutual distance of 30~cm, and the other two tiles, made of UNIPLAST scintillator, have been put in the centre of the tower. 
Signals from the four tiles have been acquired with the same digitizer used for the test-beam.

The external tiles have been instrumented with Type 2 FEE, the same used in the test-beam; their signals have been discriminated at 30 mV and put in coincidence to obtain a trigger.
The central tiles have been used for testing the four different FEE types.

\begin{figure}[hbt]
  \begin{center}
 \includegraphics[width=\textwidth]{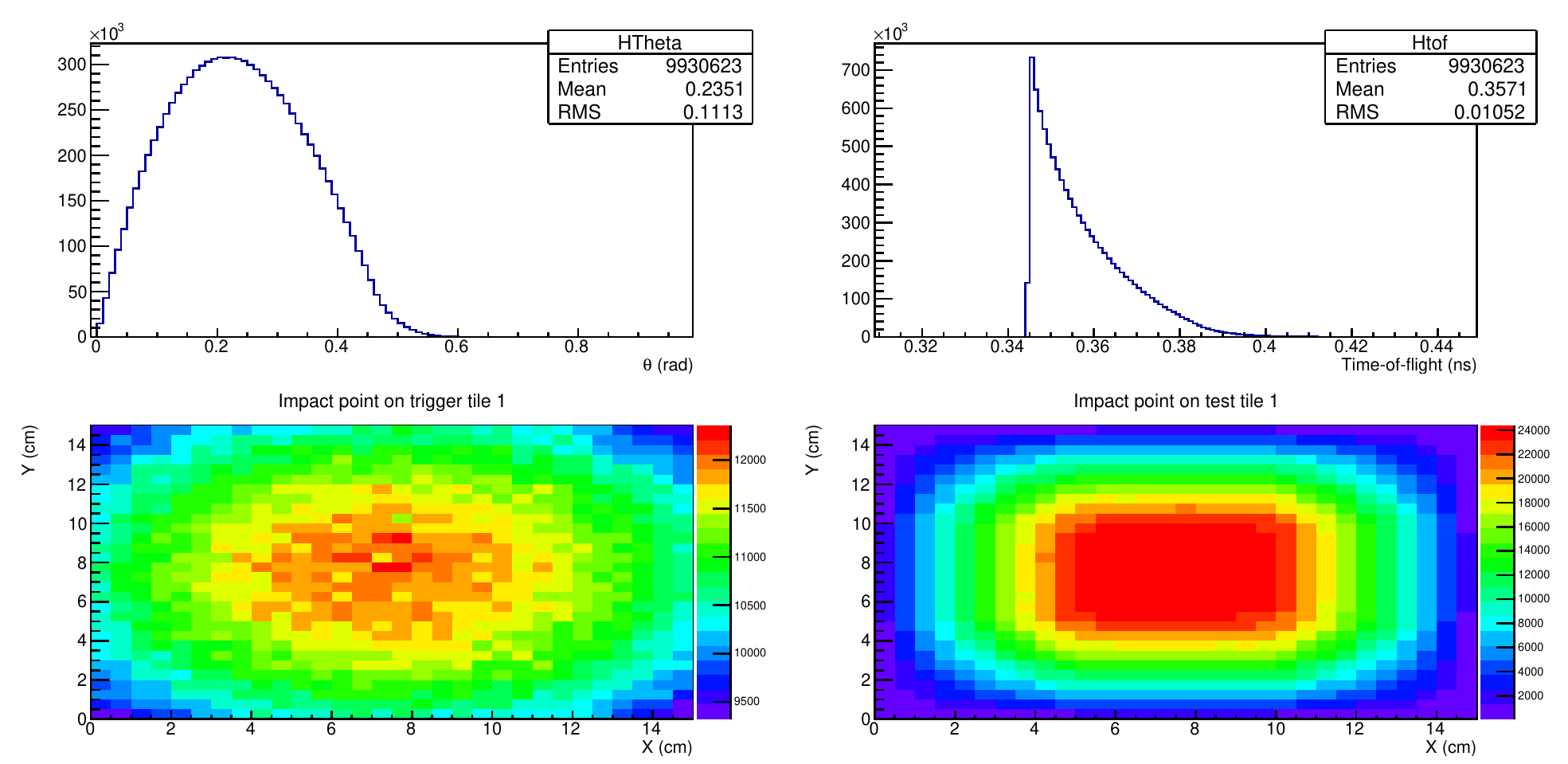}
\caption{Cosmic rays simulation: distribution of the zenith angle, $\theta$ (top left), estimated time-of-flight between the central tiles (top right), impact point spatial distribution for one of the external (bottom left) and one of the internal (bottom right) tiles.}
\label{fig:MC_cosmic}
\end{center}
\end{figure}

A toy MonteCarlo simulation has been developed to estimate systematic effects due to the cosmics angular distribution,
assuming the widely used $cos^2 \theta d\Omega$ angular distribution \cite{Zyla:2020zbs}.
The angular distribution for cosmic rays triggered by the external tiles is shown in Figure \ref{fig:MC_cosmic} together with the distributions of the estimated time-of-flight between the central ones. 
The spatial distribution of the impact point in one of the trigger and in one of the central tiles is also shown.
As a consequence of the cosmic rays angular distribution, the time-of-flight between the central tiles has an rms of 11 ps, that will be neglected in the following analysis. 
Triggered cosmic rays are focused in the centre of the inner tiles, and even though impinging on the entire surface, ten times more events crossing the central area are detected with respect to peripheral zones.
For the external tiles, used as trigger, the flux is more uniform, at the level of $\pm$20\%.

For the analysis of the digitized signals, baseline values have been estimated for each tile, averaging the first 150 samples (corresponding to 15 ns), and then subtracted.
From the study of baseline distributions, similar electronics noise values as those observed at the test beam, shown in Figure \ref{fig:baseline_study}, have been inferred.
The signals of the four tiles have been discriminated at 20\% of their amplitudes and the time resolution has been estimated using the time-of-flight between the central tiles, by fitting the distribution with a Gaussian function and dividing its sigma by $\sqrt 2$.
Events with signal amplitudes lower than 50 mV or higher than 950 mV (the digitizer has 1 V range) in either of the central tiles have been discarded.

\begin{figure}[htbt]
  \begin{center}
 \includegraphics[width=0.8\textwidth]{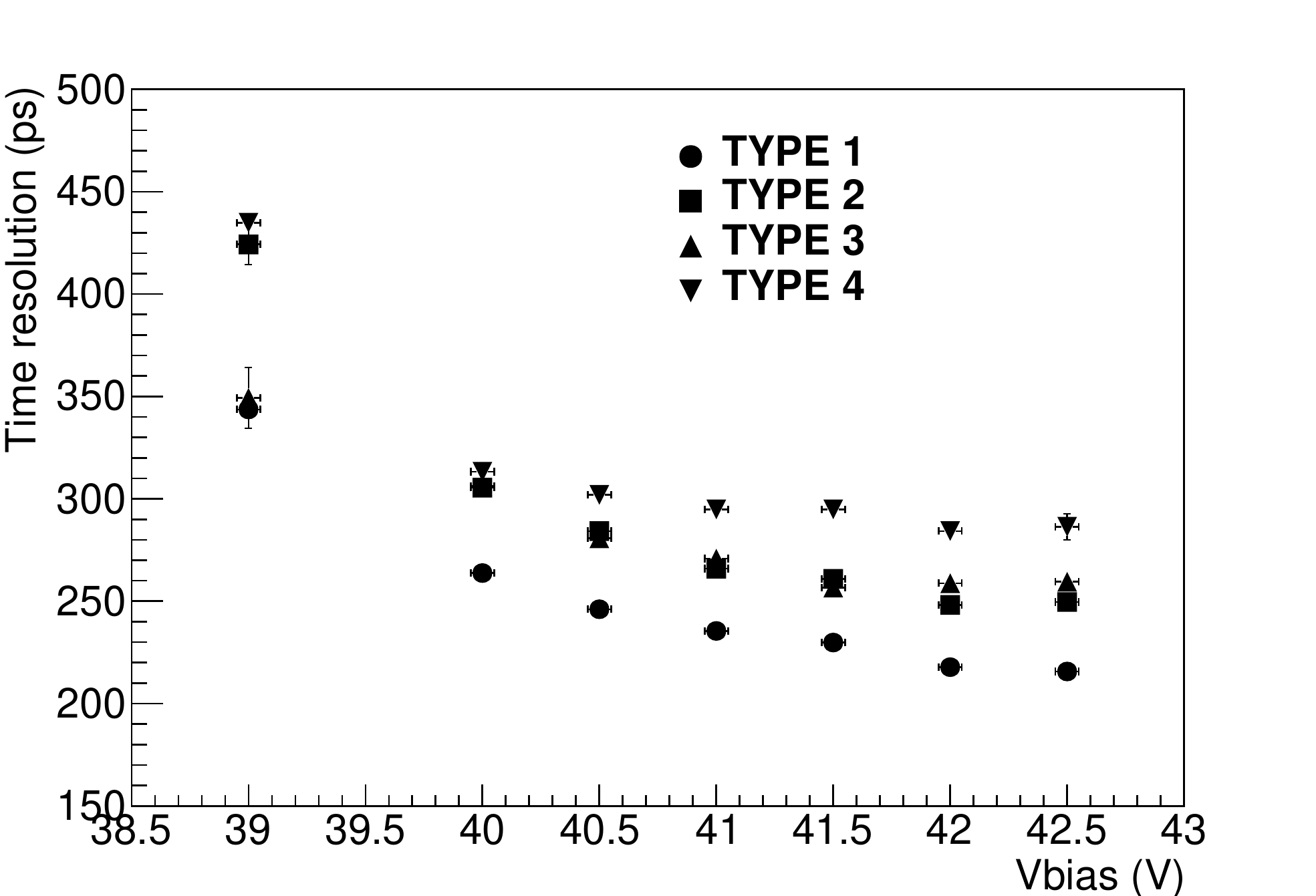}
\caption{Time resolution measured using cosmic rays with the four different electronics as a function of the SiPM bias voltage.}
\label{fig:tres_cosmic}
\end{center}
\end{figure}

The measured time resolutions with the four different FEE types are reported in Figure \ref{fig:tres_cosmic} for a bias voltage ranging from 39 V to 42.5 V at a temperature between 22 and 26$^o$C. 
The best performance is obtained with Type 1 FEE ($\sigma$ $\sim$ 230 ps for 1 V over-voltage), while the worst with Type 4; the other two types have resolutions around 260 ps.
For a better understanding, in table \ref{tab:signalchar} the amplitude Most Probable Value (MPV, obtained by means of a Landau fit) and the rise time (from 10\% to 90\% of the amplitude) are reported for the signals acquired at 41.5 V with the different electronics on tile 4: Type 1 electronics is the fastest one, but not the one with the greatest amplification.
Waveform examples acquired on tile 4 with the four different FEEs at a bias voltage of 41.5 V are shown in Figure \ref{fig:waveforms_cosmic}.
For sake of comparison, the arrival times in the tiles have been estimated also with the same algorithm used for the test beam data analysis.
The difference in time resolutions for most of the measurements is lower than 5\%, which is taken as a systematic error.

\begin{table}[hbt]
\caption{Tile 4 signal characteristics at 41.5 V bias voltage using the four different electronics types.}
\label{tab:signalchar}
\vspace{.1cm}
\begin{center}
\begin{tabular}{ccc}
\hline \hline
Electronics type & Amplitude MPV (mV) & Risetime (ns) \\ \hline
Type 1 &  281.6 $\pm$ 0.8 & 5.4 $\pm$ 0.1 \\
Type 2 &  259.8 $\pm$ 0.7  & 10.6 $\pm$ 0.1 \\
Type 3 &  170.5 $\pm$ 0.8  & 6.5 $\pm$ 0.1 \\
Type 4 &  507.7 $\pm$ 1.5  & 5.7 $\pm$ 0.1 \\ 
\hline \hline
\end{tabular}
\end{center}
\end{table}

It is worth mentioning here that the time resolutions reported in this section are expected to be worse than those measured at the test beam, as a consequence of the fact that cosmic rays are triggered on a larger area and fluctuations on the light collection time play a more important role.
To estimate the overall time resolution for an uniform illumination of the tiles under test, data have been acquired also triggering on the central tiles, put at a mutual distance of 30 cm, in order to reproduce the distribution shown at bottom left of Figure \ref{fig:MC_cosmic}.
The measured time resolution with Type 2 FEE is about 300 ps, with a 10\% systematic error from the comparison between the two employed analysis methods.
This value has to be compared with $\sim$ 220 ps measured with the focused beam at BTF facility and $\sim$ 260 ps measured with focused cosmic rays as bottom right of  Figure \ref{fig:MC_cosmic}.

Profiting of data acquired by triggering on the coincidence of the central tiles at 30 cm distance, together with those acquired for the time resolution measurements, we have compared the MPV of the signal amplitude distribution for the four different tiles at the same conditions: Type 2 FEE, $V_{bias}$=41.5 V and (almost) uniform cosmic rays impact point distribution (bottom left of Figure \ref{fig:MC_cosmic}).
The measured values are reported in table \ref{tab:tiles_cosmic}, where the same notation of the test beam for the tile identification has been used, for light yield comparison.
The best light yield is obtained for tiles 1 and 3, while worse results are obtained for tile 2 (without slots for the SiPMs) and for tile 4 (reflecting coating obtained by painting rather than etching).
Similar conclusions can be drawn from test beam data, as reported on table \ref{tab:scan_results}, where light yields can, for instance, be compared for electrons crossing the centre of the tiles (point A).

\begin{table}[htbp]
\caption{MPV of the signal amplitude distribution for the four tiles equipped with Type 2 FEE at $V_{bias}$=41.5 V.}
\label{tab:tiles_cosmic}
\vspace{.1cm}
\begin{center}
\begin{tabular}{ll}
\hline \hline
       & MPV (mV) \\ \hline
Tile 1 &  461.7 $\pm$ 2.0 \\
Tile 2 &  356.4 $\pm$ 1.2 \\
Tile 3 &  468.3 $\pm$ 1.6 \\
Tile 4 &  261.7 $\pm$ 1.1 \\ 
\hline \hline
\end{tabular}
\end{center}
\end{table}

\begin{figure}[htbt]
  \begin{center}
 \includegraphics[width=0.8\textwidth]{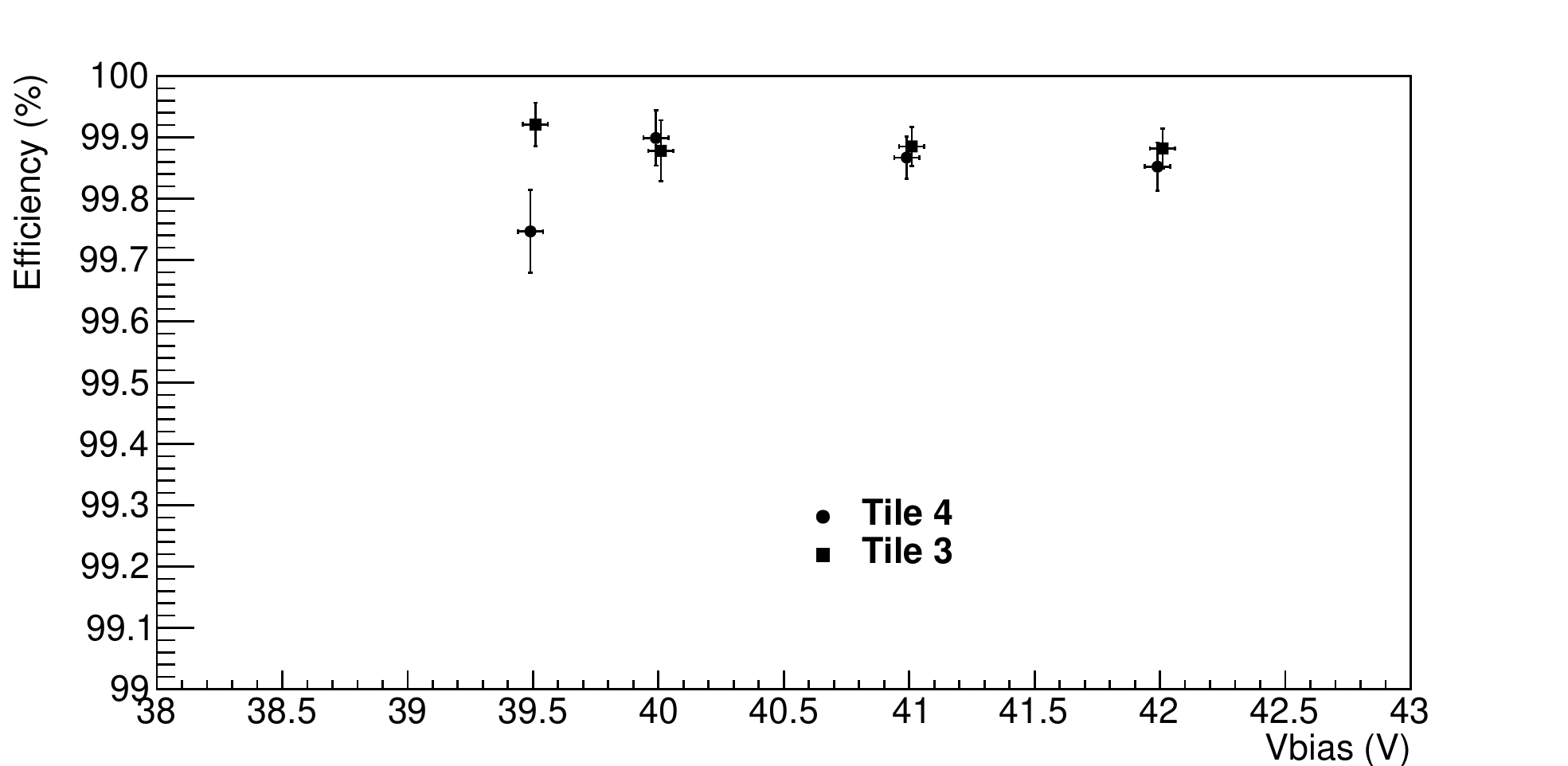}
\caption{Efficiency values measured for Tiles 3 and 4 on cosmic rays as a function of $V_{bias}$. A threshold of 30 mV on Type 2 FEE output has been applied.}
\label{fig:efficiency}
\end{center}
\end{figure}

For measuring the efficiency, two additional tiles, $(9 \times 9) \mbox{cm}^2$ wide, have been added at the two extremities of the set-up, well inside the area of the tiles under test.
Their discriminated signals have been added to the trigger in coincidence with the two external tiles.
In addition, tiles 2 and 4, as well as tiles 3 and 1, have been put in close contact each other, to have all the four tiles under test within a distance of about 10 cm.
For the efficiency estimation of tiles 3 and 4, a signal above 30 mV in a 30 ns time window with respect to the trigger is required in the other three tiles; the tile is considered efficient if a signal above 30 mV has been recorded.
In Figure \ref{fig:efficiency} the measured efficiencies as a function of $V_{bias}$ are shown for Tiles 3 and 4.
For $V_{bias}>39.5 V$ (more than 1 V over-voltage) efficiency values greater that 99.8 \% have been measured for both tiles.
Given the use of the cosmic rays without tracking devices for event reconstruction, we assume this value as a lower limit.
Similar values have been obtained for Tiles 1 and 2, exploiting the same set-up with Tiles 3 and 4 in the trigger. 

\clearpage
 \section{Simulation}

The hardware R\&D activity has been complemented by a detailed
FLUKA~\cite{fluka} simulation of the combined response of the tile and of the photo-sensors to a minimum ionizing particle. The FLUKA package (version 2020.0.6) has been used with Flair~\cite{flair} (version 2.3.0) as graphical user interface.

\begin{figure}[htbt]
  \begin{center}
 \includegraphics[width=0.4\textwidth]{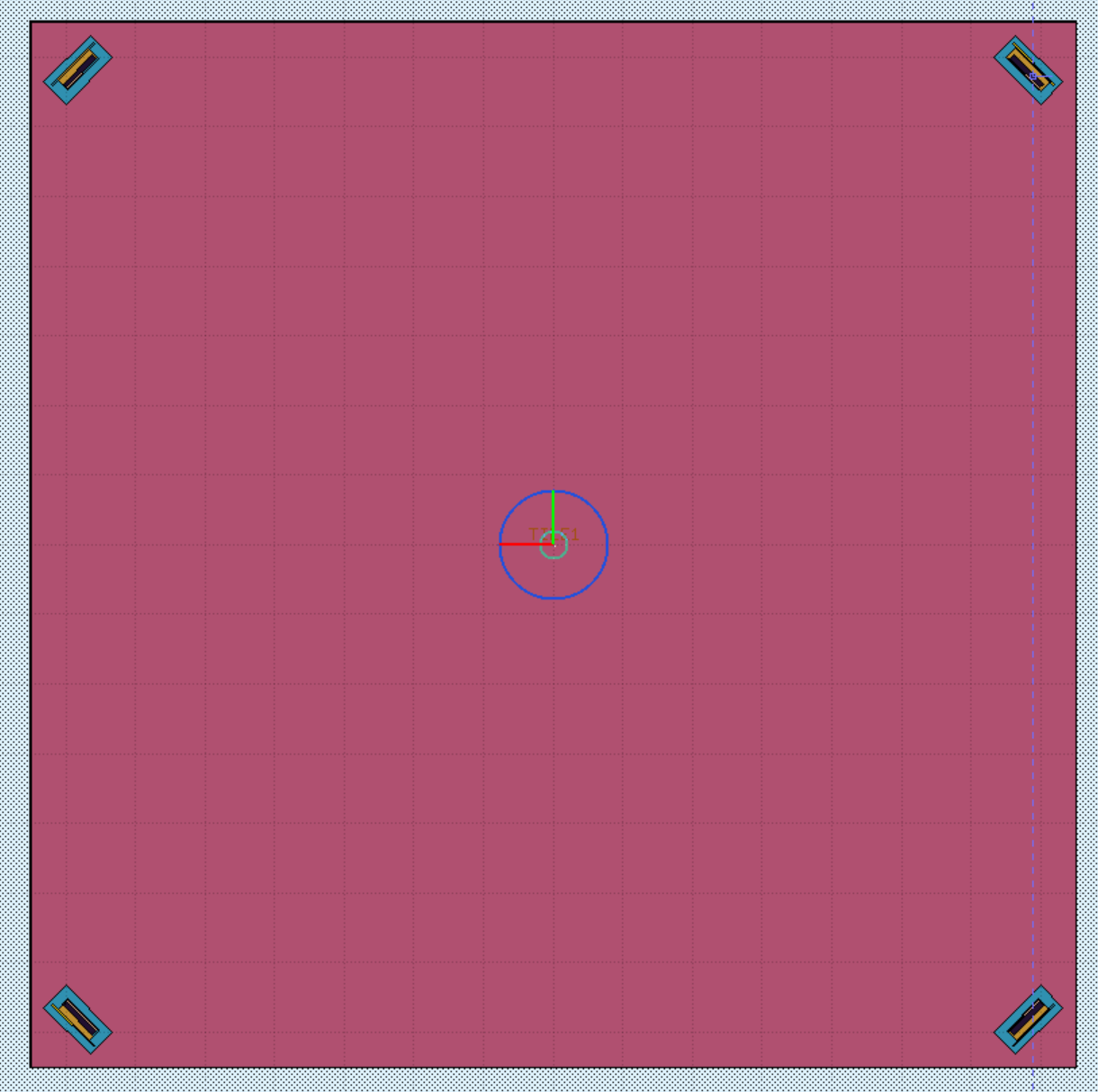} \quad \includegraphics[width=0.4\textwidth]{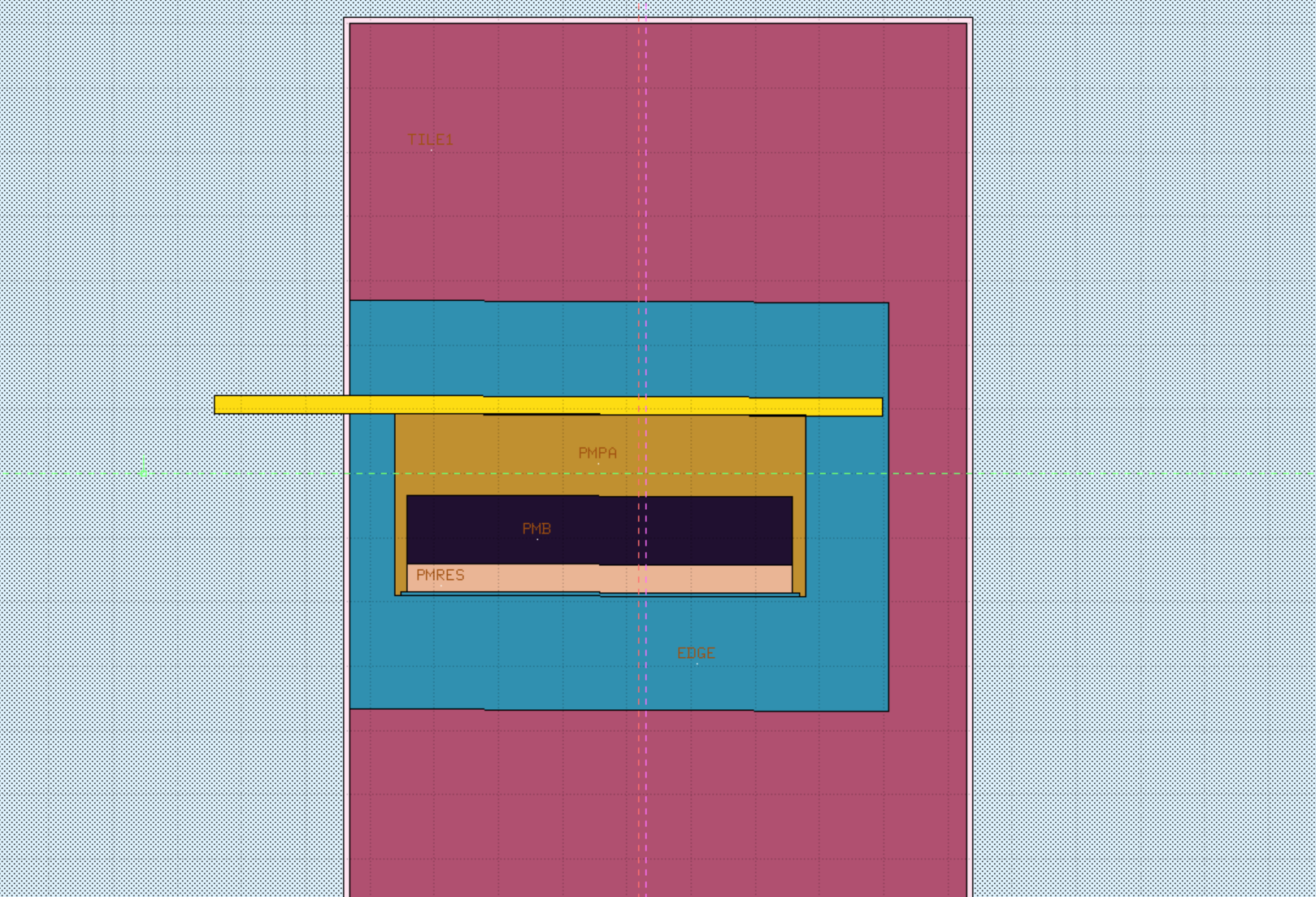}
\caption{Rendering of the tile geometry (left) used in FLUKA simulations and close-up of SiPM slot (right). }
\label{fig:sim_rendering}
\end{center}
\end{figure}

A squared ($15\times 15$) cm$^2$ tile, made of EJ200 scintillator and  read by four $(6\times 6) $mm$^2$ Hamamatsu SiPMs at the corners has been simulated. Figure \ref{fig:sim_rendering} shows a rendering of the tile geometry and a detailed view of the SiPM placed in its slot. 

The EJ200 scintillator properties (attenuation length, scintillation efficiency, emission spectrum, raise/decay times) and the photo-detection efficiency versus wavelength of the SiPMs used during tests have been implemented. 
Also the glue around the SiPMs with its optical properties has been taken into account. 
The efficiency of the optical coupling was fixed to 95\%. 
Photons hitting the $TiO_2$ coating undergo to diffuse reflection with 90\% reflectivity.
The SiPM output is calculated based on the photons reaching the SiPM window with an empirical response function: for each photoelectron an output signal is computed and the output signal waveform is produced summing over all photoelectrons.
The tile output signal is then obtained by adding all four SiPMs waveforms. 

\begin{figure}[hbt]
  \begin{center}
    \includegraphics[width=0.95\textwidth]{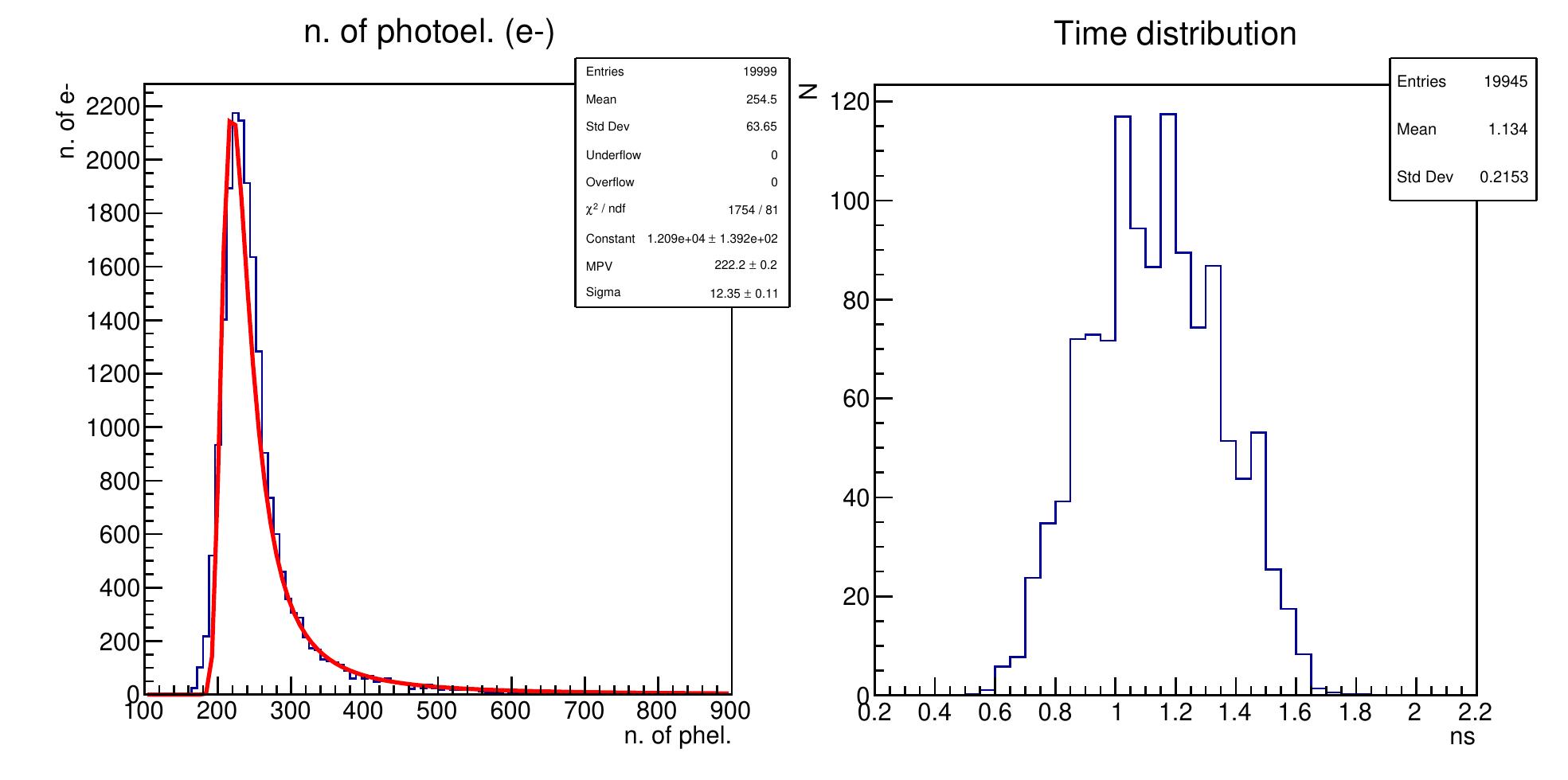}
    \caption{Distribution of the number of detected photo-electrons (left) and of the tile response time (right) with simulated electron beam events. }
\label{fig:sim_ebeam}
\end{center}
\end{figure}

We simulated the tile response to an electron beam with energy and focusing dimensions equivalent to the one used during the test beam measurement campaign, in order to perform a direct comparison. 
Simulated signals from the tiles have been discriminated at $20\%$ of their amplitudes and the corresponding times recorded.
The contribution of the time jitter due the SiPM and Type 2 FEE response has been taken into account by smearing the arrival times by a gaussian function according to the measurements reported in Section \ref{sec:time_jitter_sipm}.
The time resolution has been determined by fitting the resulting distribution with a Gaussian function. 
Figure~\ref{fig:sim_ebeam} shows the distribution of detected photons of a sample of 20k events, with the beam passing through the tile center: a MPV of $222 \pm 12$ p.e is measured,  which is in good agreement with the test beam measured value of $230 \pm 20$ p.e ( see Tab. \ref{tab:scan_results}). The simulated time resolution with a $20\%$ threshold is ~215 ps, which successfully replicates the measured value (Tab. \ref{tab:scan_results}). 

\begin{figure}[hbt]
  \begin{center}
    \includegraphics[width=0.9\textwidth]{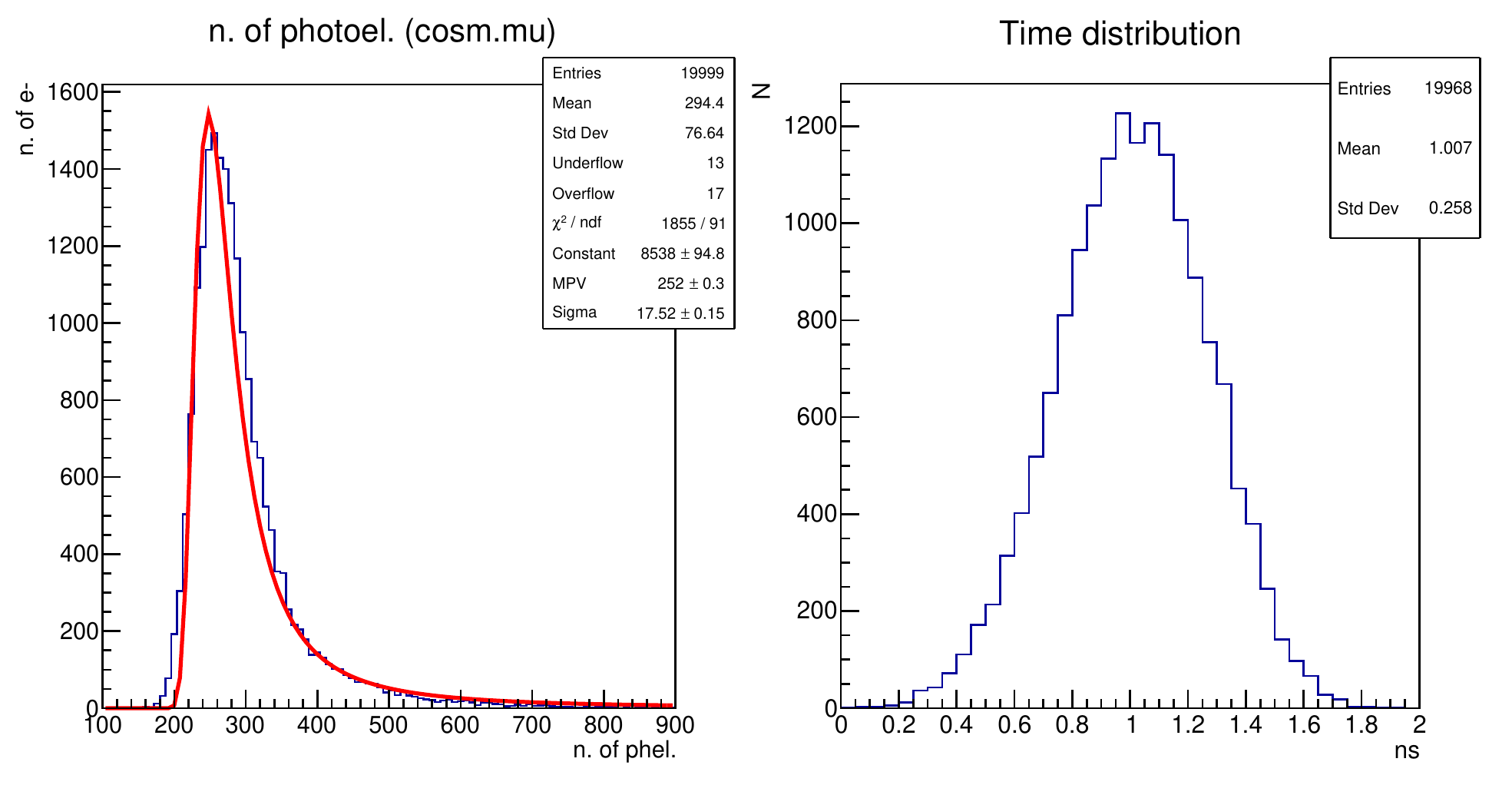} 
    \caption{Distribution of the number of detected photo-electrons (left) and of the tile response time (right) with simulated cosmic muons events.}
\label{fig:sim_cosmic}
\end{center}
\end{figure}

Similarly we simulated the tile response to cosmic muons with a $cos^2\theta d\Omega$ angular distribution as described in Section~\ref{chapter_cosmics}. 
Figure \ref{fig:sim_cosmic} shows the distribution of detected photons (left) and the time distribution (right) for a sample of 20k events generated according to the spatial distribution shown in Figure \ref{fig:MC_cosmic} (bottom right). The tile time resolution worsen with respect to the simulated electron beam data, resulting in ~258 ps, which is in agreement with the results observed in laboratory measurements with Type 2 FEE, shown in Figure \ref{fig:tres_cosmic}. 

\begin{figure}[hbt]
  \begin{center}
     \includegraphics[width=0.9\textwidth]{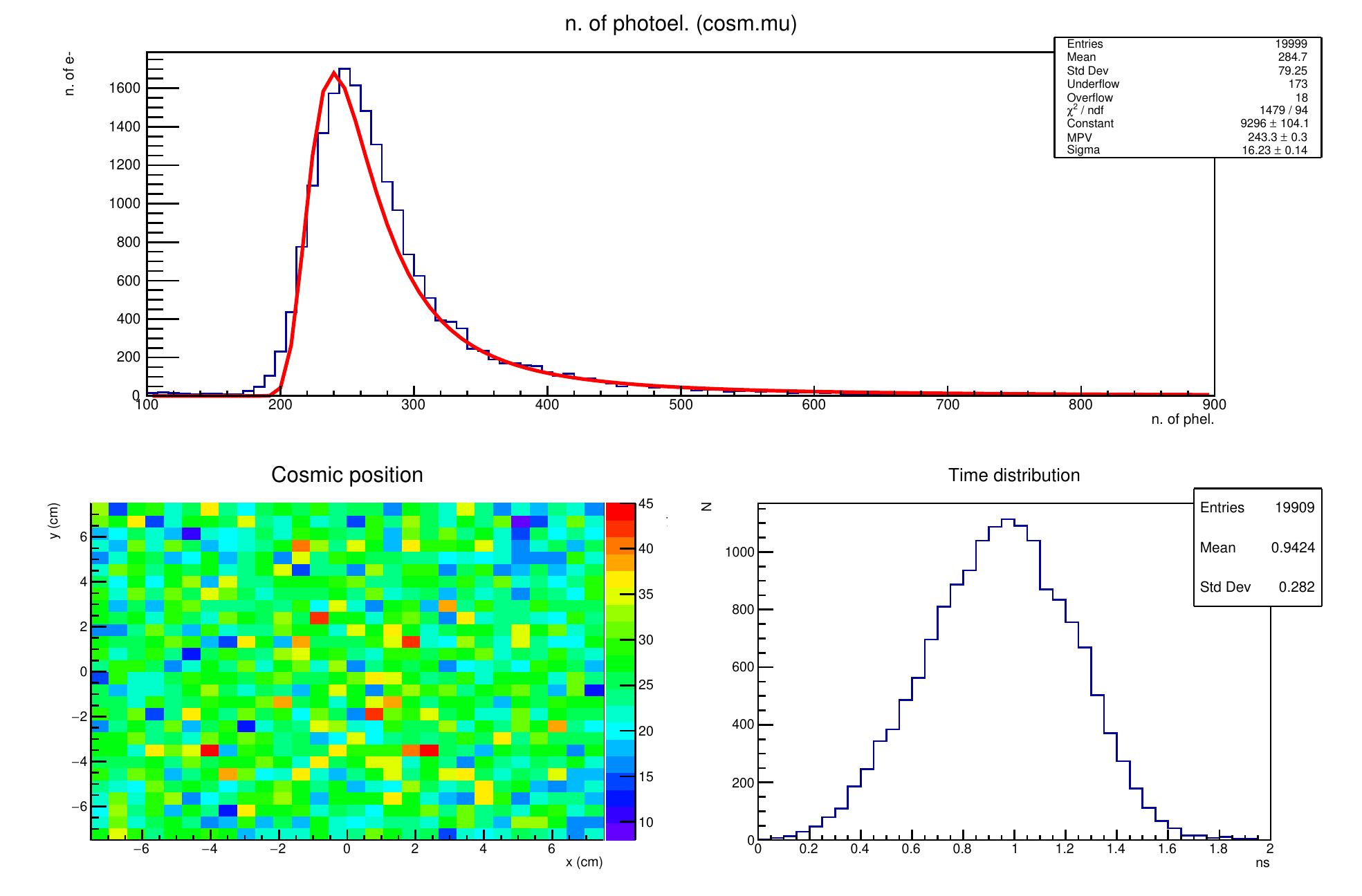}
     \caption{Tile simulation with an uniform muon flux: distribution of the number of detected photo-electrons (top), impact point spatial distribution (bottom left) and tile response time (bottom right).}
\label{fig:sim_cosmic_uniform}
\end{center}
\end{figure}

Finally, a uniform cosmic muon distribution was simulated, shown in Figure \ref{fig:sim_cosmic_uniform}. In this case the time resolution is ~282 ps, which is in slightly worse agreement with the measured value, but larger than the previously simulated distributions. 
  

 \section{Conclusions}
 \label{sec:conclusions}
 In this paper the performances of four $(150 \times 150 \times 10$)~mm$^3$ scintillating tiles, each read out by four SiPMs Hamamatsu S14160-6050HS with $(6\times 6)$~mm$^2$ active area, have been investigated in terms of light yield, time resolution and efficiency.
 Different construction techniques have also been compared.
 
 Tiles 1 (EJ200 scintillator painted with reflecting painting) and 3 (UNIPLAST scintillator with reflective layer obtained by chemical etching) have the best light yield, more than 200 photo-electrons for minimum ionizing particles impinging on the center of the tile.
 A worse performance is obtained with Tile 2 (EJ200 scintillator with SiPMs glued on cut corners rather than engraved inside the tile) and Tile 4 (UNIPLAST scintillator painted with reflective painting). 
 
 Despite the quite different light yield values, the four tiles show a similar performance in terms of time resolution, as measured in the BTF test beam.
 A possible explanation is that the dependence of the time resolution on the light yield is relaxed for a sufficiently high number (>100) of  photo-electrons.
 
 By means of cosmic muons, we have measured the overall time resolution for an uniform illumination of the tile, which is 306 ps with a systematic error of the order of 10\% from the comparison of the results with the two different considered analyses. 
 In case of a uniform illumination of the tile, a worse time resolution is expected because of the contribution due to the light propagation from the particle impact point to the nearest SiPM.
 With the cosmic ray set-up, efficiency values greater than 99.8 \% have been observed for all the tiles.
 
 The measured light yield and time resolution values have been cross-checked by means of a FLUKA based MonteCarlo simulation, which can be further used to estimate the performances of tiles with different geometries.
 The agreement within data and Montecarlo is good, at a level better than 10\% for all the data samples considered: test beam electrons as well cosmic rays, both with a focused and an uniform illuminations.
 
 Different Front-End electronics have been also tested; the best timing performances are obtained with the current conveyor configuration.
 
 This technology is therefore proven to be suitable for large area scintillating detectors when time resolution of 200-300 ps and very high efficiency are required.

\acknowledgments
We are indebted to B. Ponzio (INFN-LNF) for his support for the installation of the remote control software for the movement of the rigid frame at the BTF test beam.
We are grateful to T. Napolitano and the SPCM service of LNF for the realisation of the mechanical supports used for the tests.
Finally we acknoledge professor Y. Kudenko (INR-Moscow) for his fruitful suggestions.



\end{document}